\documentclass[prb,showpacs,amssymb,amsmath,twocolumn,superscriptaddress]{revtex4}

\usepackage{bm}
\usepackage{amsmath}
\usepackage{amssymb}
\usepackage[dvips]{graphicx}
\def\be{\begin{align}}
\def\ee{\end{align}}
\def\bea{\begin{eqnarray}}
\def\eea{\end{eqnarray}}

\newcommand{\ket}[1]{|#1\rangle}
\newcommand{\bra}[1]{\langle#1|}

\newcommand{\sgn}{\mathrm{sgn}}

\begin{document}
\title{Theory of interaction-dependent instability in quantum detection by means of Luttinger liquid tunnel junction: a rigorous theorem}

\author{Gleb A. Skorobagatko}

\affiliation{%
\mbox{%
Institute for Condensed Matter Physics of National Academy of Sciences of Ukraine,
  Svientsitskii Str.1,79011 Lviv, Ukraine%
}}

\email{ <gleb.a.skor@gmail.com>}

\date{\today}
%\pacs{}

\begin{abstract}
The low-temperature regime of charge-qubit decoherence due to its Coulomb interaction with electrons tunneling through Luttinger liquid quantum-point contact (QPC) is investigated. The study is focused on quantum detector properties of Luttinger liquid QPC. Earlier results on related problems were approximate, up to the second order in small electrostatic coupling between charge-qubit and QPC. However, here it is shown that in low-(and zero-)temperature limit the respective perturbative decoherence- and acquisition of information timescales both tend to diverge, thus, shadowing a true picture of low-temperature quantum detection for such quantum systems. Here it is shown, that one can successfully circumvent these difficulties in order to restore complete and exact picture of low-temperature decoherence and quantum detection for charge-qubit being measured by arbitrary Luttinger liquid QPC. To do this, here I prove two general mathematical statements (S-theorem and S-lemma) about exact re-exponentiation of Keldysh-contour ordered T-exponent for arbitrary  Luttinger liquid tunnel Hamiltonian.  The resulting exact formulas are believed to be important in a wide range of those Luttinger liquid problems, where real-time quantum field dynamic is crucial. As the result, decoherence- and acquisition of information time-scales as well as QPC quantum detector efficiency rate are calculated exactly and are shown to have a dramatic dependence on repulsive interaction between electrons in 1D leads of QPC. In particular, it is found that at temperatures close to zero there exists a certain well-defined threshold value $ g \approx g_{cr}(T) $ of Luttinger liquid correlation parameter $ g $ ($0 <g \leq 1 $) which serves as a sharp boundary between region of good (or even perfect) quantum detection at $ g<g_{cr} $ and the region of quantum detection breakdown for $ g>g_{cr} $. Moreover, discovered abrupt decrease of QPC quantum detector efficiency $  Q $ with the increase of $ g $ in the close vicinity of value $ g_{cr} $ represents a fingerprint of interaction-dependent \textit{instability} of all the quantum detection procedure for \textit{any} Luttinger liquid QPC quantum detector at definite low enough temperatures $ T_{cr}(g) $. The reasons behind these effects are discussed. Also, it is shown that such the low-temperature detection instability effect is able to explain a large unclear mismatch between expected and observed decoherence timescales in two recent experiments ( \textit{J.Gorman, D.G.Hasko, D.A.Williams, {\em Phys.Rev.Lett.}, {\bf 95}, 090502, (2005)} and  \textit{K.D.Petersson, J.R.Petta, H.Lu, A.C.Gossard, {\em Phys.Rev.Lett.}, {\bf 105}, 246804 (2010)} ) on charge-qubit quantum dynamics.
\end{abstract}

\maketitle

\section{Introduction}

Continuous progress in the experimentally achievable design of different electronic devices governed by quantum electron dynamics on nanoscale rises a new paradigm in the view on interactions in quantum condensed matter systems being relevant for the design of quantum computers \cite{1,2,3,4,5,6,18,19,20,21,33,34,35,36,37,38,39,40}. This results in the novel concept of \textit{quantum measurements} \cite{1,2,3,4,5,6,9,10,15,16,17,33,34,37,38,39,40}. which concerns just a measurable consequence of interaction between two subsystems of certain well-isolated quantum system \cite{1,4,10,11}. Especially, if one among two quantum subsystems (in what follows a \textit{measured system}) is prepared in a certain coherent superposition of only few well-resolved quantum states (if there are only two such states then this subsystem represents just a coherent state of a \textit{qubit}) then coupling of a measured subsystem to the (in general, many-body)  \textit{detector subsystem} (or \textit{quantum detector}) is supposed to be sensitive to the quantum state of the measured system (qubit) \cite{4,9,10,11,12,13,14,33,34,35,36}. Hence, any quantum state of a detector subsystem by definition should change during its interaction with measured quantum subsystem because of the interaction between two \cite{5,6,9,10,49}. Obviously, there are some cases (e.g. tunnel contact quantum detectors), where one is able to read out these changes in the quantum state of the detector-subsystem and then to extract from those data the information about the quantum state of a coupled measured subsystem \cite{11,12,13,14,18,19,20,21,40,41}.  However, interaction between measured- and detector subsystems obviously affect both, continuously changing also the quantum state of a measured subsystem being prepared initially \cite{5,6,21,22}.  In other words, interaction between measured- and detector- counterparts of any isolated quantum system leads to the gradual \textit{decoherence} of both \cite{5,6,7,8,37,38,39,40,41,50,54}. And this is the price for the most non-invasive measurements on quantum object one could ever perform by means of another (as small) quantum object (quantum detector)\cite{9,10,50}. 

 For the moment there exists many experimentally valuable platforms where such quantum measurement ideology can be implemented, those include measured subsystems like qubits of any kind: quantum double-dots (playing the role of charge qubits), Josephson qubits, systems of cold atoms, etc.\cite{1,2,3,4,11,12,13,14,18,19,21,33,34,40,41} - those can interact with certain quantum detector-subsystems  such as, quantum wires, quantum Hall edge states in 2D heterostructures and quantum-point contacts \cite{33,34,35,36,37}. The most straightforward practical benefits from quantum detection theory and engineering consist in the decoherence minimization in the readouts of qubit states in various quantum computation protocols including weak measurement ideology \cite{22,23,24,25,26,27,28,29} and, moreover, in the controllability of the qubit states due to the tunable interaction between quantum detector and measured quantum system\cite{30,31,32,33,34,35,36,37}. 

For instance, the described above general concept of interaction between measured system and quantum detector becomes crucial in order to perform and to understand two recent experiments on the charge-qubit state preparation, manipulation and relaxation reported in Refs.[11,12]. In both those experiments the \textit{measured} decoherence- (or equally, coherence-) timescales (being equal approximately to $ 200 ns  $ for Ref.[11] and $ 10 ns $ for Ref.[12]) were reported to be much \textit{larger} than it was initially expected for charge-qubits under consideration \cite{11,12} (e.g. in $ 10^{2} $ times larger for the experiment from Ref.[11] and $ 10^{1} $ times larger for the experiment of Ref.[12]). The authors of Refs. [11,12] refer this big variances in the measured decoherence timescales to the inevitable and uncontrollable influence of charge-qubit surroundings, however, they left this too common physical reason without its further concretization. 

In such the situation this paper can shed some light on more specific physical reasons for this experimentally observed variances in the charge-qubit decoherence timescales. The fact is, the experimentalists in both cases (of Ref.[11] and of Ref.[12]) managed to cool their quantum systems together with their environments down to extremely low operating temperatures  (indeed, they reported $ 20 mK $ in Ref.[11] and $ 80 mK $ in Ref.[12]). The reported temperatures correspond to extremely low energy scales of thermal fluctuations in these experiments (these energies are of the order of $ 10^{-5} - 10^{-6} eV $) which are much lower than operating gate voltages and are comparable only with the negligibly small effects of discontinuity in the energy spectrum of QPC electrodes (for example, the largest energy scale in QPC subsystem  is the Fermi energy of QPC electrodes and charging energy of DQD, both are typically of the order of several electron-volts \cite{11,12} ). Therefore, in this "effective" zero-temperature limit, both \textit{real}- and \textit{measured} (in the experiments of Refs.[11,12]) decoherence timescales should be referred to the only  "incoherent" effect in the system which still remains relevant in this effective zero-temperature limit\cite{5,6,10,41,50} the one is the electrostatic interaction between the electron on the DQD (i.e. charge-qubit) and electrons tunneling through QPC quantum detector while latter performs a given quantum measurement on this charge-qubit.

In the view of a leading role of Coulomb interactions in the system at low temperatures of interest, it becomes clear that a presence of long-time correlations in the quantum many-body system of QPC detector represents a single crucial factor. The latter defines both \textit{decoherence time} $ \tau_{dec} $ referred to the quantum state of a given charge-qubit and a "time of reaction" $ \tau_{acq} $ of the many-body quantum state of QPC on the simultaneous time-evolution of a charge-qubit quantum state; $ \tau_{acq} $ - is also widely known as the \textit{acquisition of information time}. \cite{5,6,10,41,50} These two time-scales are not the same. In reality, including the experiments of Refs.[11,12], the experimentalist always measures not the \textit{true} decoherence-time $ \tau_{dec} $ for a given charge-qubit quantum state but only the timescale $ \tau_{acq} $ associated with an "echo" of charge-qubit decoherence process which affects also a QPC subsystem. Remarkably, from very basic postulates about the projective measurements in quantum mechanics, one may conclude that $ \tau_{acq} \geq \tau_{dec}$ , i.e. "true" charge-qubit decoherence time never exceeds the acquisition of information timescale (otherwise, the exact result of a proper projective measurement would be known \textit{before} the corresponding wave function would "collapse" to that result and such a situation is, of course, impossible). 

That is why the ratio between these two timescales defines so-called \textit{quantum detector efficiency rate}: $ Q= \tau_{dec}/\tau_{acq} \leq 1$ where the case: $ Q=1 $ (i.e. when $ \tau_{dec}=\tau_{acq} $ and the same quantum processes of interaction are responsible both for decoherence of measured system and for its feedback to detector) is known in the literature as \textit{quantum limit of detection} which is typically realized only in the absence of large thermal fluctuations in quantum system of interest \cite{10,50}. However, as we will see below, in the case of definite weak electron-electron interactions in QPC quantum detector, even at zero- and near-zero temperatures, the acquisition of information time $ \tau_{acq} $ can be still much longer than a "true" decoherence time $ \tau_{dec} $ for a given charge-qubit interacting electrostatically with given QPC quantum detector. This, in turn, can result in the "breakdown" of all the quantum detection procedure in the zero-temperature limit for QPC detectors those having a proper electron-electron interection in their leads in order to provide a situation: $ Q \rightarrow 0 $ . Hence, this quantum detection "breakdown" effect if to be revealed for a qiven QPC at zero- or near-zero temperature could perfectly explain a big variance in the magnitudes of characteristic decoherence timescales being measured in the experiments of Refs.[11,12]. Here I shall describe the latter effect in details.  

In general, due to one-dimensional effective geometry of QPC the problem of related electron-electron interactions becomes very important and respective quantum dynamics of such systems has many parallels with physics of non-equilibrium phenomena in 1D quantum mesoscopic systems\cite{7,8,15,45,47}. Especially, for the moment there exists a number of both theoretical\cite{5,6,7,8,15,20,38,41} and experimental\cite{12,16,32,33,34} studies on different aspects of non-equilibrium quantum dynamics (including some simplest cases of decoherence \cite{41}) in mesoscopic quantum systems due to electron-electron interaction between their counterparts. However, the attempts to bring together electron-electron interactions in the leads of QPC \cite{7,8,15,16,17,45,46,47} and continuous quantum measurement approach \cite{5,6,41} within a quantum detector ideology\cite{9,10} have been missed in the literature till recent time. A first detailed study on dynamics of QPC quantum detector with arbitrary electron-electron interaction in its 1D Luttinger liquid electrodes has been performed by the author and his colleagues\cite{50} for the regime of weak electron tunneling in the finite temperature limiting case. The main achievement of the latter work \cite{50} is a generalization of the entire quantum detector concept to the case of arbitrary electron-electron interaction in the quantum detector subsystem (at non-zero temperatures and arbitrary bias voltages). As general result, it was found  that strong electron-electron interactions shift quantum limit of detection to much lower temperatures and much higher bias voltages \cite{50}. 

Nevertheless, the properties of a described quantum detection procedure in the system: charge-qubit + tunnel junction of two 1D Luttinger liquid quantum wires have remained unclear at temperatures near the absolute zero. Physical reason for that is straightforward: at near-zero- (and zero) temperatures quantum fluctuations in Luttinger liquid tunnel junction become large with respect to thermal noise and, as the result, respective tunneling time - diverges already in its first perturbative order\cite{39,41,45,46,47,50,53,55} due to Kane-Fisher effect\cite{47}. The latter means actually that an infinite number of virtual charge-qubit-assisted resonant tunnelings of interacting electrons - contributes transport characteristics of Luttinger liquid tunnel junction of interest. And one needs to sum up the infinite number of all contributions from all the orders in qubit-QPC interaction just to obtain the exact expression for both decoherence- and acquisition of information time-scales in the zero- (and near-zero)-temperature limit of the problem. This implies a necessity to know the exact expression for respective generating function for arbitrary interacting Luttinger liquid tunnel junctions since up to the moment such exact expressions were known only for the case of non-interacting electrons in QPC electrodes (i.e. only for Fermi-liquid leads)\cite{48,49}. In what follows this theoretical problem will be solved explicitly in its most general Luttinger liquid realization.  

\begin{figure}
\includegraphics[height=13 cm,width=9 cm]{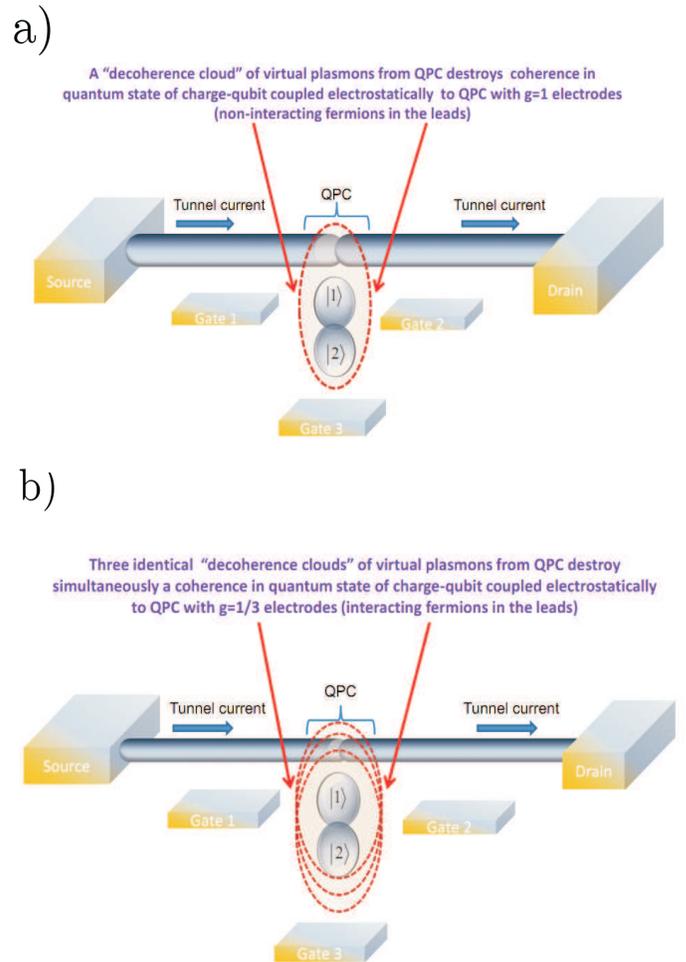}
\caption{\textit{Schematic picture of the model setup similar to ones have been used in the experiments of Refs.[11,12]: charge-qubit (double quantum dot or DQD with one excess electron on it) electrostatically interacting with biased QPC (which serves as the current-carrying quantum detector and a source of decoherence for the former) and three controlling gates (which independently modulate charge-qubit evolution). Fig.1a depicts the most common case of 3D quantum wires  with non-interacting electrons at $ g=1 $ (Fermi liquid leads)in the role of QPC electrodes, while on Fig.1b one can see Luttinger liquid realization of the same setup with 1D quantum wires in the role of QPC leads of interacting electrons with $ g=1/3 $. It will be shown in the text that low-temperature decoherence and quantum detection in systems of such type are governed by quantum states of DQD excess electron "dressed" into Kondo-like polaronic "clouds" of virtual plasmons from QPC. Semi-transparent red dashed ellipses around the DQD and the QPC tunnel contact on Figs.1a,b correspond to such "dressing": these ellipses depict different numbers of polaronic "Kondo-like clouds" (or equally, "decoherence clouds") of virtual plasmons in different cases of charge fractionalization in QPC electrodes: $ g=1 $ (one Kondo-like "cloud" on Fig.1,a) and $ g=1/3 $(three identical Kondo-like "clouds" on Fig.1,b). } 
}
\end{figure}

Especially, below I will prove an important theorem about the exactness of the re-exponentiation procedure for definite types of averages from \textit{T-}exponent with non-linear operator-valued function of bosonic quantum field in its power. This mathematical result (being useful by itself for many problems of Luttinger liquid real-time dynamics out-of-the equilibrium) seems to be one among very few known examples of exact analytic formulas for averaged real-time evolution operators with highly non-linear functions of bosonic quantum field in the Keldysh-contour-ordered \textit{T-}exponents. For instance, these exact results can be implemented within full-counting statistics ideology\cite{48,49,52} as well as for Rabi-oscillating quantum systems\cite{11,12,13,14,52}.   

On the other hand, formulas obtained below provide a desired exact description of quantum detection and decoherence in its zero-temperature limit at arbitrary bias voltages for arbitrary electrostatic interaction between charge-qubit and its quantum detector. As the consequence, here it will be shown that decoherence mechanism near the zero-temperature limit is governed by the processes which are analogous to a well-known Kondo physics in the 1D Anderson impurity model \cite{55,56,57} (see also Fig.1). It also turns out, that strong electron-electron interactions in the leads of quantum detector can sufficiently "improve" quantum QPC-detector efficiency while electron-electron interactions of small or intermediate strength always only suppress it. 

Moreover, it will be found that there exists a fixed coupling- and temperature-dependent threshold value for the electron-electron interaction in the leads, corresponding to a fixed critical value $ g_{cr} $ of Luttinger liquid correlation parameter in the QPC quantum detector electrodes. Remarkably, it will be shown that for $ g < g_{cr}$ (strong electron-electron interactions in the leads) one has: $ Q \lesssim 1$ thus acquisition of information time being measured by means of the statistics of charge transfer through  Luttinger liquid QPC - represents also "true" decoherence time for the measured charge-qubit. Whereas, in the case of QPC electrodes with: $  g_{cr}< g \leq 1$ (i.e. for QPC electrodes with moderate or weak electron-electron interactions) one has a situation where $ Q \rightarrow 0$  and measured acquisition of information time $ \tau_{acq} $ has nothing to do with real decoherence time of given charge-qubit (since in the latter case: $ \tau_{acq} \gg \tau_{dec} $). In addition, the obtained "steepness" of QPC detector efficiency in the  vicinity of a "threshold" value $ g_{cr} $  - represents a clear fingerprint of \textit{instability} in the quality of quantum detection procedure for any QPC quantum detectors with $ g \simeq g_{cr} $. Remarkably, it also will be shown, that such a \textit{quantum detection instability} effect might be responsible for the observed variances in the decoherence time-scales measured in the experiments of Refs.[11,12].  Qualitative physical reasons for such an instability are also discussed. Obviously, the results of this paper can potentially contribute to qualitative explanations of decoherence phenomena for many other actual mesoscopic systems in the zero-temperature limit, e.g. those involving double-quantum dots and 1D tunnel jucntions. As well, these results can be implemented to a proper engineering in a variety of prospective experiments with charge-qubits interacting with quantum wires.

This paper has a following structure: In Section 2) I describe the underlying theoretical model together with its brief experimental justification (while closer relation of the model to the actual experiments is discussed separately in the Appendix A). In Section 3) I represent some basic tools for the problem: a general formula for quantum detector efficiency rate for QPC quantum detector \cite{50} and the orthogonality catastrophe pre-factor as well as the formulation of low-temperature uncertainty in QPC quantum detector efficiency. In Section 4) I formulate basic S-theorem and S-lemma, both needed to obtain the consequent physical results, and discuss the physical reasons behind the validity of these rigorous mathematical statements. In Section 5) using the results of Sec.2)-4) I derive central equations of this paper: the exact formula for the efficiency of Luttinger liquid QPC quantum detector at zero- and near-zero temperatures and the analytic estimation for the low-temperature threshold value $ g_{cr} $ of Luttinger liquid interaction parameter in QPC electrodes. In Section 6) I explain the qualitative physics behind the revealed novel effects and make conclusions. The relation of the theoretical model under consideration to actual experimental background is discussed in details in Appendix A, while lengthy and cumbersome proofs of S-theorem and S-lemma one can find in Appendix B to this paper.

\section{Model}

As long as this paper is targeted to the qualitative interpretation of the real experiments \cite{11,12}, it is worth to justify the underlying theoretical model by comparing it with real quantum systems from Refs.[11,12]. It is done in details in the Appendix A to this paper (whereas a reader interested only in theoretical aspects of this research could safely omit Appendix A). From Fig.1 one can extract a minimal needed information about the geometry and constituent details of the setup (Fig.1a schematically depicts the QPC quantum detector with "bulk" Fermi liquid (FL) electrodes, while Fig.1b refers to the "interacting" case of relatively narrow 1D Luttinger liquid (TLL-)quantum wires in the role of QPC electrodes). 

From all the above as well as from Refs.[11,12] and arguments of Appendix A, it becomes clear, that a QPC quantum detector in our case represents a biased tunnel contact which connects right (R) and left (L) semi-infinite Luttinger liquids in the regime of weak tunneling\cite{46,47,50}. The quantum state of a double-dot system (-DQD or charge qubit) being electrostatically coupled to QPC  affects tunneling of the electrons through the latter\cite{10,41,50}(see Fig.1). It is also assumed everywhere in what follows that there are no any strong magnetic fields in the system, all leading effects have electrostatic nature, hence, one can model this situation by spinless electrons (as it takes place in the experiments of Refs.[11,12]).

The total Hamiltonian of our problem, thus, should consist of three terms: 

\begin{equation}
H_{\Sigma} = H_{LL} + H_{DQD}+ H_{int},
\end{equation}

where $H_{LL}$ represents the Hamiltonian of the left and right Luttinger liquids, $H_{DQD}$ that of the DQD and $H_{int}$ stands for the interaction between these two parts of the system\cite{50}(see also Appendix A for related details and justification of the model for real experimental setups). If we consider the QPC to be located at $x=0$, then (Here and everywhere in all the numbered formulas I put $ \hbar=1 $ and $ e=1 $ though in the text I, sometimes, restore dimensional units for clarity)

\begin{equation}
H_{LL} =\frac{1}{ 2 \pi} \sum_{j = L, R} v_g  \int^0_{- \infty} \left\{ g \left(\partial_x \varphi_j \right)^2 + \frac{1}{g} \left( \partial_x \theta_j  \right)^2 \right\} \rm{d}x ,
\end{equation}
 
where $\theta_{L (R)}(x)=\pi\int_{-\infty}^{x}dx' \rho_{L ( R )}(x')$ and $\varphi_{L( R)}(x)=\pi\int_{-\infty}^{x}dx' j_{L ( R )}(x')$ are the usual charge- and phase- bosonic quantum fields in the Luttinger liquid description of semi-infinite 1D quantum wire, those corresponding to fluctuating parts of charge- and current electron densities in the QPC leads \cite{46,47,50,53}: $ \rho_{L( R )}(x)= \sum_{c=1,2}:\Psi_{c,L(R)}^{\dagger} \Psi_{c,L(R)}(x):$, $ j_{L( R )}(x)= \sum_{c=1,2}(-1)^{c}:\Psi_{c,L(R)}^{\dagger}\Psi_{c,L(R)}(x): $, where fermionic creation (annihilation) field operators: $ \Psi_{c,L(R)}^{\dagger}$ and $ \left(  \Psi_{c,L(R)}(x) \right)  $ (with $ c =1,2$)- create (annihilate) a left-(with $ c=1 $) and right-moving (with $ c=2 $) chiral fermions at the point $ x \leq 0$ of either left ($ L $) or right ($ R $) 1D electrode of a QPC (here $ :..:  $ stands for normal ordering of respective fermionic operators). In Eq.(2) $g$ is a dimensionless correlation parameter which is defined as $ g \approx \left( 1 + U_{s}/2E_{F} \right) ^{-\frac{1}{2}}$ ( where $ U_{s} $ is the potential energy of short-range Coulomb interaction in QPC elctrodes, for repulsive interactions it fulfils $0 < g \leqslant 1$)  while $v_g$ is the group velocity of collective plasmonic excitations in the leads\cite{46,47,50,53}. Notice, that we have chosen the coordinate systems on both left $ L $ and right $ R $ electrodes of QPC in such a way that $x$ increases from $-\infty$ to the point $ x=0 $, where the QPC is located\cite{47,50,53}. Here it is also presumed (without loss of generality) that all electron energies in the system are counted from the Fermi level of QPC electrodes, provided that for biased tunnel junction $ \mu_{L}=eV $ and $ \mu_{R}=E_{F}=0 $ (where $ V $ is corresponding bias voltage, see e.g. Ref.[50] and Appendix A). 

As for double-quantum dot(DQD) subsystem, its role in the total Hamiltonian of the system is restricted by only two lowest unoccupied molecular (LUMO) levels of DQD with one excess electron shared between these two levels, those forming a two-level electron quantum system, or simpler, a charge-qubit, since, importantly, these two (LUMO)levels correspond to different potential wells (or equally to different quantum dots )of DQD "droplet" (for further details on charge-qubit preparation see Appendix A ). Thus, a quantum state of one excess electron in the resulting two-level system is completely defined by three energy parameters (see also Appendix A): the energies of two corresponding levels $ \varepsilon_{1,2} $ (each level is localized in the respective quantum dot of DQD structure) and tunnel coupling  $ \gamma $ between two quantum dots in the DQD system), the latter coupling is non-zero due to the possibility of electron tunneling between two potential wells (otherwise one could not prepare initially the required coherent state of the electron on DQD subsystem). From the above it follows that $ \vert \varepsilon_{1} - \varepsilon_{2} \vert < E_{c,1,2} $ (here $ E_{c,1,2} $ is a Coulomb charging energy for respective quantum dot, see details in Appendix A). These three energy parameters of a charge-qubit can be regulated (nearly independently from each other) by means of three (or more) external gates coupled capacitively to the DQD (see Refs.[11,12] and Fig.1).
Thus, the Hamiltonian of the DQD reads

\begin{equation}
H_{QD} = \sum_{n = 1, 2} \varepsilon_n c_n^\dagger c_n + \gamma \left(c_1^\dagger c_2 + c_2^\dagger c_1 \right) , 
\end{equation}

where $c_n^\dagger \left( c_n \right)$ are fermionic operators of creation-(destruction) of an electron in the n-th quantum dot ($n = 1, 2$), and $\varepsilon_n$ are electron energies in DQD (with respect to the Fermi energy level of the TLL leads which is chosen in the above to be equal to zero) , $\gamma$ represents the inter-dot tunnel coupling constant. 

 This enables experimentalist to modulate the time-evolution of charge-qubit state-vector on the Bloch sphere, thus, preparing and manipulating coherently different charge-qubit quantum states of DQD subsystem \cite{11,12}. For example, one can easily tune (independently from each other) two energy levels $ \varepsilon_{1,2} $  to their "resonant" position  where $ \varepsilon_{1}=\varepsilon_{2}=E_{F} $, however, just because the latter condition, which fixes the values of all three gate voltages, it is not guaranteed that in such a resonant case $ \gamma $ will be also equal to zero, rather one will always have a natural situation where $ \gamma \neq 0$ (this is exactly the case in Refs.[11,12]). Especially, in the resonant situation $ \gamma $ is the only parameter which defines the time-evolution of charge-qubit, being proportional to the frequency of respective Rabi oscillations $ \omega_{R}=\gamma / \hbar $ when DQD subsystem is decoupled from its QPC quantum detector (see e.g. Refs.[11,12]).  However, one can easily engineer the setup where $  \omega_{R}^{-1} \gg \tau_{dec},\tau_{acq} $ provided that in this case $ \gamma $  is one of the smallest energy scales in the system: $ \gamma \ll T,eV \ll \Lambda_{g} $ (where $ \Lambda_{g} $ is a high-energy cutoff of the model, see text below and in the Appendix A) which is an equivalent of the limit $ \gamma \rightarrow 0 $. To achieve the limit $ \gamma \rightarrow 0 $ where Rabi oscillations of charge-qubit are "frozen out" under the resonance condition modulated by the gates, one can either couple capacitively to the DQD a one more independent gate (which is not the case here) or, alternatively, prepare a very fine-tuned asymmetric DQD-"droplet" providing a very small tunneling between its two consistuent quantum dots. The latter situation is considered below (see also a scetch of DQD on Figs.1a,b) while another important limit being interesting for applications  $ T,eV \ll \gamma < \Lambda_{g}  $ is beyond the scope of this paper. 
 
 Here it is worth to emphasize, that despite the fact in the experiments of Refs.[11,12] one has one more energy scale: $ \gamma=\hbar \omega_{R} \gtrsim T,eV $ (which results in the situation where $ \omega_{R}  \gg \tau_{acq}^{-1}$) measured in these experiments acquisition of information time-scales still can be qualitatively interpreted using the theoretical results of this paper being obtained in the limit $ \gamma \rightarrow 0$. This is because the DQD decoherence- and its "feedback" to QPC quantum detector are both defined by the electrostatic interaction between two conserving (though not static) electric charges (i.e. between one electron tunneling through QPC and Rabi-oscillating state of one extra electron on DQD). Since Rabi oscillations in the DQD subsystem are unable to change the total electric charge of DQD they also cannot change the qualitative picture of decoherence and acquisiton of information defined by a total DQD charge. Thus, the only possible effect from non-zero $ \gamma=\hbar \omega_{R} $ is a certain modification of a bias voltage in all formulas obtained below\cite{52}. And as one can see from the derivation of all formulas in below, such a modification does not affect the character of mathematical functions (e.g. critical exponents) and, hence, the nature of physical effects have been obtained in what follows in the limit: $ \gamma \rightarrow 0 $ and being responsible for what happens in the system in the low-temperature limit.  

This limit together with the above resonant condition: $ \varepsilon_{1}=\varepsilon_{2}=0 $ justifies an important simplifications of our model: $ H_{DQD}=0 $ - actually, this is the same approach as one have been used in our previous paper on quantum detection of charge-qubit by means of TLL quantum detector at non-zero temperatures\cite{50}.  
 
Taking into account that within the "weak tunneling" approach for the TLL tunnel junction model\cite{47,50}  "charge"- bosonic fields $ \theta_{L,R}(x=0,t) $ are pinned on the edges of respective QPC electrodes at point $ x=0 $ by means of the condition\cite{46,47,50,53}  $ \theta_{L}(x=0,t)= \theta_{R}(x=0,t)=0$  and introducing following non-local bosonic charge- and phase-fields\cite{50} $ \theta_{\pm}=\left[\theta_L \pm \theta_R\right] $ and $ \varphi_{\pm}=\left[ \varphi_L\pm \varphi_R \right] $ one can write down the interaction Hamiltonian $ H_{int} $ for the underlying model as follows (for justification one can see Appendix A and Refs.[47,50])
 \begin{align}\label{eq:H_int}
  H_{int} = \sum_{n = 1, 2}
  [\lambda_n \partial_x \theta_+ + \tilde{\lambda}_n \cos \left(
  \varphi_-  + \rm{eV} t \right)]|_{x = 0} c_n^\dagger c_n\,,
\end{align}
where $\lambda_n$ represents electrostatic coupling between one-electron charge density on the $ n $-th quantum dot of DQD and the charge density on the edges of Luttinger liquid electrodes which model given QPC, while $\tilde{\lambda}_n=t_n/(\pi a_0)$, with $t_n$ being a "bare" tunneling amplitude for a given QPC. Both quantities $\lambda_n$, $ \tilde{\lambda}_n $ ($ n=1,2 $)  therefore depend on the state of the electron on DQD (i.e. on the quantum state of a given charge-qubit) \cite{41,50}.  The parameter $a_0$ is the lattice constant of the model. This constant provides a natural high-energy cut-off, and goes to zero in the continuum limit\cite{53}. Notice, that in the limit of weak tunneling which concerns us here, the difference between electrostatic potentials of the leads can be treated as the local voltage drop at the impurity site of corresponded lattice model\cite{46}. Quantity $\rm{eV}$ refers to this drop being proportional to the external bias voltage applied between left and right Luttinger liquid electrodes of QPC. 

As the result, if one "initialize" our charge qubit (i.e. the DQD) in its quantum coherent state $\ket{\phi_0}=\alpha\ket{1}+\beta\ket{2}$ at $ t=0 $ then the subsequent time-evolution of the qubit will be governed only by its interaction with coupled QPC quantum detector. This can be represented in terms of the reduced density matrix $\tilde{\rho}$ for the entire quantum system, where environment (i.e. detector) degrees of freedom have been traced out \cite{41,50}. If the initial state of the entire quantum system at $ t=0 $ is a product state $\ket{\phi_0}\otimes\ket{\chi(0)}$ where $\chi(t)$ is the ground state of $H_{LL}$ Hamiltonian at time $t$, then, in the interaction representation\cite{41,50} with respect to $ H_{int} $, one can write for  $\tilde{\rho}$ following expression 
\begin{align}\label{eq:rho_t}
 \tilde{\rho}_{mn}(t)=\rho_{mn}(0)\bra{\chi(t)}\overline{\mathcal{U}}_n(t)\mathcal{U}_m(t)\ket{\chi(t)}
\end{align}
with $\rho(0)=\ket{\phi_0}\bra{\phi_0}$, $m,n=1,2$, $H_{QD}\ket{n}=\varepsilon_n\ket{n}$ and $  \mathcal{U}_n \left( t\right)= \mathcal{T}_t \exp \left\{ - i \int^t_0 d \tau \tilde{H}^{\left( n \right)}_{{int}}  \left( \tau \right) \right\} $, $ \overline{\mathcal{U}}_n \left( t\right)= \overline{\mathcal{T}}_t \exp \left\{ i \int^t_0 d \tau \tilde{H}^{\left( n \right)}_{{int}}  \left( \tau \right) \right\} $, where $\mathcal{T}_t$, $\overline{\mathcal{T}}_t$ denote time- and anti-time -ordered exponents\cite{50} and $\tilde{H}^n_{int}=\bra{n}H_{int}\ket{n}$ with $ H_{int} $ from Eq.(4). Then the environment-induced decoherence will be encapsulated in the off-diagonal matrix elements of  $\tilde{\rho}$ \cite{41,50}.   As it has been already shown in Ref.[50], since bosonic fields $\varphi_+$ and $\varphi_-$ commute with each other at arbitrary times $ \left[\varphi_+(t),\varphi_-(t')\right]=0 $ it is straightforward to evaluate their vacuum expectation values separately. Then the off-diagonal elements $\tilde{\rho}_{mn}$ with $m\neq n$ of the reduced density matrix will factorize exactly on two time-dependent averages  
\begin{align}\label{eq:rho_mn}
\tilde{\rho}_{mn}(t)=\rho_{mn}(0)Z_{(mn)}(t)\tilde{Z}_{(mn)}(t).
\end{align}
For the orthogonality catastrophe contribution in the Luttinger liquid case one has \cite{50,54}
\begin{eqnarray}
\nonumber
Z_{(12)}(t)=\left\langle \exp{ \left\{i\frac{g\left(\lambda_{1}-\lambda_{2} \right)}{v_g}\left[\varphi_+(t)-\varphi_+(0)\right]\right\}}\right\rangle \\
=\left [ \frac{\pi T/\Lambda_{g}}{\sinh (\pi T\cdot t)}\right ]^{2g(\Delta\lambda/\Lambda_{g})^{2}}
\end{eqnarray}
(where $ \Delta\lambda=\lambda_{1}-\lambda_{2} $ and $\Lambda_{g}=\Lambda_{0}/g  $ , $ \Lambda_{0}\simeq E_{F} $ is a high-energy cut-off in the noninteracting (Fermi-liquid) case $ g=1 $ ). Remarkably, formulas (6,7) represent exact result \cite{50,54}. Whereas for the tunneling contribution to reduced density matrix one has following general expression\cite{50}
\begin{equation}
\tilde{Z}_{(mn)}(t)=\left\langle \mathcal{T}_K\exp(-i\int_{\mathcal{C}_K}\eta_{mn}(\tau)A_{0}(\tau){\rm d}\tau)\right\rangle .
\end{equation}
Here the integral is taken along the complex Keldysh contour $ \mathcal{C}_K\in (0-i\beta;t) $ with $ \eta_{mn}(\tau)=\pm \vert \tilde{\lambda}_{n}-\tilde{\lambda}_{m} \vert $ ($ m,n=1,2 $) on the upper (lower) branch of Keldysh contour, whereas  "quantum potential" $ A_{0}(\tau) $ is defined as $ A_{0}(\tau)=\cos{\left[\varphi_-(\tau)+eV\tau\right]} $ ( see Refs.[49,50]). Calculation of both contributions from Eqs.(7,8) should provide one with all the necessary information about the charge-qubit total decoherence rate in our model \cite{50}. However, in all the preceding literature on related problems for any propagators similar to $ \tilde{Z}_{(mn)}(t) $ there were no attempts to proceed calculations beyond the second order in small tunnel coupling $ \tilde{\lambda}_{m} $ (see e.g. Refs.[8,41,48,49]). This is because all the higher orders are usually irrelevant for the most of electron transport characteristics (such as, for example, average current and shot-noise power) while the difficulties  of corresponding explicit real-time Keldysh contour calculations do increase very fast for each  subsequent order in $ \tilde{\lambda}_{m} $. As well, the same mathematical difficulties one faces calculating time-dependent generating function $ W(\Delta \tilde{\lambda},t) $ within the full-counting statistics formalism for any quantum-point contact \cite{48,50}. This is why in the related preceding papers\cite{48} as well as in our recent paper \cite{50} on quantum detector characteristics all calculations which involve $ \tilde{\lambda}_{m} $ were performed perturbatively, only up to the second order in small tunnel couplings, which is relevant under condition $ \Delta \tilde{\lambda}=\vert \tilde{\lambda}_{1}- \tilde{\lambda}_{2} \vert /\Lambda_{g} \ll 1 $.  
However, below it will be shown that for the characteristics of QPC quantum detector near zero temperature existing perturbative results are not enough while corresponding mathematical difficulties can be completely resolved leading to the exact statements about the low-(and zero-)temperature quantum detection.

\section{Quantum detector efficiency at low temperatures}

In our preceding work\cite{50} it was shown that at finite temperatures a quantum detector efficiency ratio $ Q_{\Sigma} $ for our model is defined by the following formula $ Q_{\Sigma}=\tau_{dec}/\tau_{acq}=\frac{1}{a_{s} \eta}\left\lbrace 1 - \sqrt{1-\left[a_{s}\tanh(eV/2T)\right]^{2} }\right\rbrace / \left\lbrace 1+(\Gamma_{ort}/\Gamma_{t}) \right\rbrace $, which though is valid only up to the second order in small $ \Delta \tilde{\lambda} $. Here quantities $ \Gamma_{ort} $ and $ \Gamma_{t} $ represents the orthogonality catastrophe- and tunneling contributions to the total decoherence rate being calculated in Ref.[50] only up to the second order of perturbation theory in small coupling constant $ \Delta \tilde{\lambda} $ and $ a_{s}=2\eta/(1+ \eta^{2}) $ and $ \eta=(\tilde{\lambda}_{1}-\tilde{\lambda}_{2})/(\tilde{\lambda}_{1}+ \tilde{\lambda}_{2}) $ -characterize the asymmetry of QPC coupling to each among the charge-qubit quantum states. But since quantity $ \Gamma_{ort} $ makes sense only asymptotically (just because the "orthogonality catastrophe" coefficient is just the pre-exponential factor in the off-diagonal density matrix element as it can be seen from Eq.(7)) this fact automatically implies that the latter perturbative formula makes sense only for $ t \gg 1/\pi T $ (i.e. then $ T\neq 0 $).

Nevertheless, taking this problem formally one can write down expression for our QPC-detector efficiency at near-zero temperatures (i.e. when for $ t \rightarrow \infty $ one has $ t < 1/\pi T $, meaning that $ T \rightarrow 0$). The result is
\begin{equation}
\lim_{T \rightarrow 0} Q_ {\Sigma}=Q=\frac{W(\Delta \tilde{\lambda},T)}{\left\lbrace \Gamma(\Delta \tilde{\lambda},T)+\tilde{\Gamma}_{K}(\Delta \lambda) \right\rbrace},
\end{equation} 
here I put for simplicity $ a_{s},\eta \approx 1 $ - in the case of highly asymmetric coupling of charge qubit states to QPC detector (this is the situation most suitable for quantum detection). Obviously, in Eq.(9) $ W(\Delta \tilde{\lambda},T) $ should refer to exact expression for the acquisition of information rate at zero temperature, while: $ \Gamma(\Delta \tilde{\lambda},T) $ and $ \tilde{\Gamma}_{K}(\Delta \lambda)  $ should stand for exact decoherence rates due to electron tunneling and due to orthogonality catastrophe, respectively. As it follows from the exactness of the expression (7) for the orthogonality catastrophe pre-factor, one can easily make an explicit estimation for the value of $ \Gamma_{ort}(\Delta \lambda, T \rightarrow 0)=\tilde{\Gamma}_{K}(\Delta \lambda) $ (here we have $ t < 1/\pi T $, $ T \rightarrow 0$). Indeed, one may put $ \tilde{\Gamma}_{K}(\Delta \lambda)=\hbar /\tau_{ort,0} $, where $ \tau_{ort,0} $  is the characteristic time after which under condition $ t < 1/\pi T $ the pre-factor $ Z_{mn}(t) $ decreases in $ e^{1} $ times. Then from Eq.(7) one immediately obtains
\begin{equation}
\tilde{\Gamma}_{K}(\Delta \lambda)=\frac{\Lambda_{g}}{\pi}\exp \left[ -\frac{1}{2g}\left ( \frac{\Lambda_{g}}{\Delta\lambda}\right )^{2}\right ]. 
\end{equation} 
Interestingly, formula (10) reproduces well-known expression for the Kondo temperature or, equally, the formula for the width of so-called Kondo-resonance \cite{55,56} with the appropriate Kondo-coupling constant $ J_{K} \approx U^{2}/\varepsilon_{F} $ being replaced by $ \Delta \lambda^{2}/\Lambda_{g} $, while quantity $ 2g/\Lambda_{g}\approx 2g^{2}/\varepsilon_{F} $ plays the role of the edge density of states $ \nu $ in the related Anderson impurity problem \cite{55}. The latter remarkable correspondence is because in our case electrostatically coupled charge-qubit plays the role of a localized impurity in the Anderson model which, in turn, can be mapped onto the Kondo problem \cite{55}. As it has been already mentioned in the above, the appropriate estimation for both $ W(\Delta \tilde{\lambda},T) $ and $ \Gamma(\Delta \tilde{\lambda},T) $ at near-zero temperatures is a much more difficult task. However, formally, one can always write down these quantities as infinite Taylor power series in the dimensionless coupling strength $ \left( \Delta \tilde{\lambda}/\Lambda_{g} \right)^{2} $. Notice, that in the real low-temperature experiments with charge-qubits\cite{11,12} , the latter coupling is not necessary small because, in according with estimation has been made below in the Appendix A , one has for this parameter $ \left( \Delta \tilde{\lambda}/\Lambda_{g} \right)^{2} \simeq \left( gE_{c,n}/E_{F}\right)^{2} $  and since the Coulomb charging energy $ E_{c,n} $ of DQD is of the order of Fermi energy $ E_{F} $ in QPC electrodes (especially under the "resonance" condition for charge-qubit: $ \varepsilon_{1}=\varepsilon_{2}=E_{F} $), see Appendix A for details), one can write approximately: $\left( \Delta \tilde{\lambda}/\Lambda_{g} \right)^{2} \simeq g^{2} $, obviously this quantity is of the order of $ 1 $ then $ g $ is close to one( for example for $ 0.9 < g < 1  $). The general form of these series reads
\begin{eqnarray}
W(\Delta \tilde{\lambda},T)=\left( \dfrac{\Delta \tilde{\lambda}}{\Lambda_{g}} \right)^{2}w_{1}(T)+\left( \dfrac{\Delta \tilde{\lambda}}{\Lambda_{g}} \right)^{4}w_{2}(T)+\ldots 
\end{eqnarray}
and
\begin{eqnarray}
\Gamma(\Delta \tilde{\lambda},T)=\left( \dfrac{\Delta \tilde{\lambda}}{\Lambda_{g}} \right)^{2}f_{1}(T)+\left( \dfrac{\Delta \tilde{\lambda}}{\Lambda_{g}} \right)^{4}f_{2}(T)+\ldots
\end{eqnarray}
The usual practice here is to take into account only the first term in both expansions from Eqs.(11,12) being most relevant for the main electron transport characteristics at non-zero temperatures (such as average current and shot-noise) and, at the same time, those one can most easy calculate perturbatively, taking into account only the lowest order in the expansion of Eqs.(11,12). This gives us the result\cite{45,46,47,50}  $ w_{1}(T),f_{1}(T) \rightarrow 0 $ at $ eV, T \rightarrow 0 $. However, exactly because of this result and, as well, because $ \tilde{\Gamma}_{K}(\Delta \lambda) $ -is also exponentially small (see Eq.(10)), one needs to know all the other orders in Eqs.(11,12) in order to reconstruct correct quantum detector efficiency behaviour (9) at zero and near-zero temperatures. 

\section{S-Theorem and S-Lemma}

Below the Summation(or S-) theorem and S-lemma  will be formulated and proved (see corresponded detailed proofs in the Appendix B). These rigorous mathematical statements both tell us that \textit{two infinite sums from Eqs.(11,12) converge exactly to two corresponding exponential functions with their powers being equal only to first terms in the expansions of Eqs.(11,12)}, providing the exactness of the related re-exponentiation procedure\cite{50} in the calculation of the reduced density matrix of Eq.(6). The S-theorem for function $ \Gamma(\Delta \tilde{\lambda},T) $ reads (its generalization for the function $ W(\Delta \tilde{\lambda},T) $ is analogous) 

$ \lozenge $ \textit{Summation(or S-)Theorem}: 
$\blacktriangle $ \textit{For any exponential bosonic operator of the form}
\begin{align}
 A_{0}(\tau)=\cos{\left[\varphi_{-}(\tau)+f(\tau)\right]}.
 \nonumber
\end{align}
\textit{where $ f(\tau) $ can be any function of time which fulfils condition  $ f_{+}(\tau) = f_{-}(\tau) $, with $ f_{\pm}(\tau) $ being its values on the upper and lower branches of Keldysh contour in the complex plane and $ \varphi_{-}(\tau) $ is time-dependent bosonic field with zero mean $ \langle \varphi_{-}(\tau)\rangle = 0 $, which fulfils commutation relation of the form}
\begin{align}
\begin{split}
\left[\varphi_{-}(\tau_{n}),\varphi_{-}(\tau_{n'})\right]=-2i\vartheta_{g}\sgn{\left[\tau_{n}-\tau_{n'}\right]}
 \nonumber
\end{split}
 \end{align}
 \textit{where $ \vartheta_{g}=const $ (notice, that in our particular case\cite{50,53} $ \vartheta_{g}=\pi/g $)}

$ \blacktriangledown $ \textit{it follows for the average (taken over bosonic ground state  of corresponded "free" TLL Hamiltonian ) for the time-dependent Keldysh contour-ordered exponential }
\begin{align}
\begin{split}
\tilde{Z}_{12(21)}(t)=\left\langle \mathcal{T}_K\exp(-i\int_{\mathcal{C}_K}\eta_{12(21)}(\tau)A_{0}(\tau){\rm d}\tau)\right\rangle \\
=\langle \overline{\mathcal{T}}_t \exp{\left\{i\tilde{\lambda}_{1(2)}\int_{0}^{t} {\rm d}\tau\cos{\left[\varphi_-(\tau)+f(\tau)\right]} \right\}}\times\\
\mathcal{T}_t \exp{ \left\{-i\tilde{\lambda}_{2(1)}\int_{0}^{t} {\rm d}\tau\cos{\left[\varphi_-(\tau)+f(\tau)\right]} \right\}}\rangle \\
=\exp \left\lbrace - \textit{F}_{12(21)}(t) \right\rbrace 
\end{split}
\end{align} 
\textit{where for the function $ \textit{F}_{12}(t)=\textit{F}^{\ast}_{21}(t) $ one has}
\begin{align}
\begin{split}
\textit{F}_{12(21)}(t)=\left( \tilde{\lambda}_{1(2)}-\tilde{\lambda}_{2(1)}\right)\left[ \tilde{\lambda}_{1(2)}e^{i \vartheta_{g}}-\tilde{\lambda}_{2(1)}e^{-i \vartheta_{g}} \right]\\
\times \left\lbrace \int \int_{\textit{C}_{K}} d \tau_{1} d \tau_{2}\langle A_{0}(\tau_{1}) A_{0}(\tau_{2})\rangle_{S} \right\rbrace \\
=\left(\tilde{\lambda}_{1(2)}-\tilde{\lambda}_{2(1)}\right)\left[ \tilde{\lambda}_{1(2)}e^{i \vartheta_{g}}-\tilde{\lambda}_{2(1)}e^{-i \vartheta_{g}} \right]\\
\times \left\lbrace 2\int^{t}_{0} d \tau_{1}\int^{t}_{0} d \tau_{2}\langle A_{0}(\tau_{1}) A_{0}(\tau_{2})\rangle_{S} \right\rbrace.  
\end{split}
\end{align} 
\textit{Here the complex Keldysh contour is denoted as $ \mathcal{C}_K\in (0-i\beta;t) $ and "quantum" field $ \eta_{12(21)}(\tau)=\pm \vert \tilde{\lambda}_{1}-\tilde{\lambda}_{2} \vert $ - on the upper (lower) branch of Keldysh contour; the averages of the type $ \langle A_{0}(\tau_{1})A_{0}(\tau_{2})\rangle_{S} $  are "symmetrized" with respect to permutation of time arguments $ \tau_{1}\leftrightarrow \tau_{2}$  and}
\begin{align}
\begin{split}
\langle A_{0}(\tau_{l+1}) A_{0}(\tau_{l})\rangle_{S}=u(\tau_{l+1}-\tau_{l})\cdot F(\tau_{l+1}-\tau_{l})\\
=e^{\frac{-\left\langle \left[\varphi_{-}(\tau_{l+1})-\varphi_{-}(\tau_{l}) \right]^2\right\rangle}{2}}\cdot F(\tau_{l+1}-\tau_{l})
\end{split}
 \end{align}  
\textit{where}
\begin{align}
\begin{split}
 \nonumber
F(\tau_{l+1}-\tau_{l})=F(\tau_{l}-\tau_{l+1})= \cos{[f(\tau_{l})-f(\tau_{l+1})]}
\end{split}
 \end{align} 
 \textit{-represents definite symmetric function with respect to permutation of its two time arguments $ (\tau_{l}$ and $\tau_{l+1})$ . %which depends on $ \tau_{l} $ and $ \tau_{l+1} $ only by means of functions  $ f(\tau_{l}) $ and  $ f(\tau_{l+1}) $.%
 }   $ \blacksquare $
 
The proof of this theorem, in turn, makes use of the following \textit{Summation (or S-)lemma} which is also proven here in the Appendix B. S-lemma reads

$ \lozenge $ \textit{S-Lemma}: 

$\blacktriangle $ \textit{For any exponential bosonic operator of the form}
\begin{align}
 A_{0}(\tau)=\cos{\left[\varphi_{-}(\tau)+f(\tau)\right]}.
 \nonumber
\end{align}
\textit{where $ f(\tau) $ can be any function of time which fulfils condition  $ f_{+}(\tau) = f_{-}(\tau) $, with $ f_{\pm}(\tau) $ being its values on the upper and lower branches of Keldysh contour in the complex plane and $ \varphi_{-}(\tau) $ is time-dependent bosonic field with zero mean  $ \langle \varphi_{-}(\tau)\rangle = 0 $, which fulfils commutation relation of the form}
\begin{align}
\begin{split}
\left[\varphi_{-}(\tau_{n}),\varphi_{-}(\tau_{n'})\right]=-2i\vartheta_{g}\sgn{\left[\tau_{n}-\tau_{n'}\right]}
 \nonumber
\end{split}
 \end{align}
 \textit{where $ \vartheta_{g}=const $ (notice, that in our particular case\cite{50,53}) $ \vartheta_{g}=\pi/g $ }

$ \blacktriangledown $ \textit{it follows for the symmetrized averages (with respect to free TLL-bosonic Hamiltonian ground state) of $ n $-th order in $  A_{0}(\tau) ($ with $ n=2m $ - even integer number)}
\begin{align}
\begin{split}
\int^{t}_{0} d \tau_{1}\int^{\tau_{1}}_{0} d \tau_{2}\ldots \int^{\tau_{k-2}}_{0} d \tau_{k-1}\int^{\tau_{k-1}}_{0} d \tau_{k}\times  \\
 \int^{t}_{0} d \tau'_{1}\int^{\tau'_{1}}_{0} d \tau'_{2}\ldots \int^{\tau'_{(n-k)-2}}_{0} d \tau'_{n-k-1}\int^{\tau'_{(n-k)-1}}_{0} d \tau'_{n-k}\\
\times \langle \mathcal{T}_K A_{0}(\tau_{k})\ldots A_{0}(\tau_{1}) A_{0}(\tau'_{n-k})\ldots A_{0}(\tau'_{1})\rangle_{S} \\
= \frac{1}{(n/2)!} \prod_{l=1}^{n/2} \left\lbrace 2\int^{t}_{0} d \tau_{l+1}\int^{t}_{0} d \tau_{l}\langle A_{0}(\tau_{l+1}) A_{0}(\tau_{l})\rangle_{S}\right\rbrace . 
\end{split}
\end{align} 
\textit{where any average $ \langle A_{0}(\tau_{k})\ldots A_{0}(\tau_{1})A_{0}(\tau'_{n-k})\ldots A_{0}(\tau'_{1})\rangle_{S} $ is consisted of $ n $ ($ n $ is even) operators $ A_{0}(\tau_{r})$ and is symmetrized with respect to permutation of its time arguments $ \tau_{r}\leftrightarrow \tau_{r+m}$ in any pair $ \tau_{r}, \tau_{r+m}$ ($ r,m=1..n $) and}
\begin{align}
\begin{split}
\langle A_{0}(\tau_{l+1}) A_{0}(\tau_{l})\rangle_{S}=u(\tau_{l+1}-\tau_{l})\cdot F(\tau_{l+1}-\tau_{l})\\
=e^{\frac{-\left\langle \left[\varphi_{-}(\tau_{l+1})-\varphi_{-}(\tau_{l}) \right]^2\right\rangle}{2}}\cdot F(\tau_{l+1}-\tau_{l})
\nonumber
\end{split}
 \end{align}  
\textit{where}
\begin{align}
\begin{split}
 \nonumber
F(\tau_{l+1}-\tau_{l})=F(\tau_{l}-\tau_{l+1})= \cos{[f(\tau_{l})-f(\tau_{l+1})]}
\end{split}
 \end{align} 
 \textit{is a definite symmetric function of $ (\tau_{l},\tau_{l+1})$. %which depends on $ \tau_{l} $ and $ \tau_{l+1} $ only by means of functions  $ f(\tau_{l}) $ and  $ f(\tau_{l+1}) $.%
 }  $ \blacksquare $

The above S-theorem justifies the validity and the \textit{exactness} of the re-exponentiation procedure being applied to bosonic average of the form of the l.h.s. of Eq.(13) and gives the result of such re-exponentiation in the compact analytic form of the r.h.s. of Eq.(13) and Eq.(14) with known pair correlator from Eq.(15). (The latter correlator has been already calculated earlier in the Ref.[50].) At the same time, the S-lemma  (related to S-theorem, but more "wide")- justifies the \textit{exactness} of the factorization of Wick-theorem expansion in the important case of exponential operators, those representing time-dependent bosonic fields with "step-function"-valued c-number commutators, if the "vacuum" averages from these fields have zero mean values (as the result of normal ordering) - as it take place for the fluctuating Luttinger liquid bosonic phase-fields under consideration. Obviously, one may also check the validity of S-theorem (and S-Lemma) by direct calculation: just by comparing the convergence of a summation over any finite number of terms in the power expansion of Eq.(8) to the r.h.s. of Eq.(13) (the latter can be done by means of numerical calculation). Practically, the validity of S-theorem means, that easiest way to obtain all the \textit{exact} formulas below is to keep in the power expansion of Eq.(8) only the terms of the second order in $ \tilde{\lambda}_{1(2)} $ -and then just to re-exponentiate them. In such the case S-Theorem guarantees that all resulting formulas will be exact sum of \textit{all} the infinite number of orders in $ \tilde{\lambda}_{1(2)} $. 

Here I would like to comment on this remarkable result briefly. Mathematically, the validity of S-Theorem (and S-Lemma)in the Luttinger liquid model of tunnel junction in weak tunneling regime means the \textit{exact cancellation of all "crossing" diagram contributions in all time integrations}. The latter fact seems to be quite natural physically, since all these "crossing diagrams" in the real-time domain describe "non-time ordered" fluctuations of bosonic phase-fields which are significant only on short time-scales of the order $ \lesssim \Lambda_{g}^{-1} $, where $  \Lambda_{g} $ - is a high-energy cut-off for a given Luttinger liquid Hamiltonian. In other words, one might say, that the whole effect from all the "crossing" diagrams of this model - consists only in the renormalization of plasmon group velocity in given low-energy 1D "free" TLL Hamiltonian $ H_{LL} $ which describes QPC electrodes. Hence, in the weak tunneling regime under consideration all relevant effect from these "crossing" diagrams is already taken into account by means of mathematical formulation of a given quadratic Luttinger liquid Hamiltonian $ H_{LL} $ with definite value of Luttinger liquid correlation parameter $ g $. Another similar consequence from the S-Theorem being valid for the weak tunneling regime under consideration (i.e. when bosonic charge-fields are pinned on the edges of QPC electrodes\cite{46,47,50} by means of "weak tunneling" boundary conditions: $ \theta_{\pm}(x=0,t)=0 $) states that above exact re-exponentiation should automatically account for all possible virtual electron tunnelings through such a tunnel junction. Therefore, such phenomena as "Mahan-exciton" formation\cite{57} (in the vicinity of QPC tunnel contact) does not take place in our system (or more precisely, the processes of such type occur (in the case of weak tunneling ) on too short timescales of the order of $ \lesssim \Lambda_{g}^{-1} $ so, respective nonequilibrium effects should average to zero\cite{47} within Luttinger liquid model framework. S-lemma and S-theorem, in fact, show how that proceeds mathematically. 

\section{Low-temperature limit for QPC quantum detector efficiency and quantum detection instability }

Obviously both S-theorem and S-lemma enable one to derive the exact formula for quantum detector efficiency   in the limit of zero and near-zero temperature. Especially, taking into account Eqs.(9-16) one can easily obtain
\begin{equation}
Q=\frac{1}{a_{s} \eta} \dfrac{\left\lbrace 1 - \sqrt{1-\left[a_{s}\tanh(eV/2T)\right]^{2} }\right\rbrace}{\left\lbrace 1+ \left( \dfrac{\Lambda_{g}}{\Delta\tilde{\lambda}}\right)^{2}\dfrac{\tilde{\Gamma}_{K}(\Delta\lambda)}{f_{1}(T)} \right\rbrace}  
\end{equation}
where $ \tilde{\Gamma}_{K}(\Delta\lambda) $ is from Eq.(10) while $ \left(\dfrac{\Lambda_{g}}{\Delta\tilde{\lambda}}\right)^{2}\dfrac{1}{f_{1}(T)} $ represents \textit{exact} time of decoherence of charge qubit due to applied bias voltage with $ f_{1}(T) $ being explicitly calculated earlier in the Ref.[50]  as the first-order term in the corresponded infinite perturbative series (see e.g. previous section and Appendix B) 
\begin{equation}
f_{1}(T)=\dfrac{\Lambda_{g}}{4}\left( \dfrac{2\pi T}{\Lambda_{g}}\right)^{2/g - 1}\dfrac{\vert \Gamma\left(\dfrac{1}{g} + i \dfrac{eV}{2\pi T} \right)\vert^{2} }{\Gamma\left( \dfrac{2}{g}\right) } \cosh \left( \dfrac{eV}{2T}\right).
\end{equation}

Now, having in hands \textit{exact} Eqs.(10,17,18) we are able to analyse all the physics near the zero temperature in our system. On Fig.2 the exact quantum detector efficiency $ Q $ of Eqs.(17,18) is plotted as the function of Luttinger liquid correlation parameter $ g $ in the leads of QPC, for different bias voltages in the low temperature limit. The plots of Fig.2(a,b) demonstrate the role of electron-electron interactions in the low-temperature quantum detection. Especially, while the effect of bias voltage is quite predictable: the higher the driven voltage - the better the detection is \cite{50}; the role of electron-electron interactions in the electrodes of QPC in zero-temperature quantum detection procedure is much more dramatic. On can see from Figs.2a,b that at $ T \rightarrow 0$ in the system there exists a certain  threshold value $ g_{cr} $ of Luttinger liquid correlation parameter, which separates regions of good (and even perfect) quantum detection at any bias voltage when $  g<g_{cr}$  from the region of the detector breakdown at $ g>g_{cr} $ where the detector efficiency $ Q $ is close to zero (see Fig.2b). Moreover, from Fig.2,b one can conclude that derivative $\partial Q/\partial g $ from the low-temperature detector efficiency of Eq.(17) taken with respect to the TLL-correlation parameter $ g $ in the leads of QPC - even tends to diverge at certain "critical" value $ g_{cr} $ of correlation parameter in the limit $ T\rightarrow 0 $. Notice, however, the latter interesting effect in its most distinct realizations (like one on Fig.2,b) should be referred only to the case of extremely low temperatures in the model (of the order of $ 10^{-4} mK $ as one can see on Fig.2,b) which of course one could hardly realize in any contemporary cryogenic experiment\cite{11,12}. Thus, the case of most abrupt slopes (such as one on Fig.2,b) might be just the artefact of the application of Luttinger liquid (TLL)description beyond its low-energy limit, e.g. at temperatures $ T < \hbar v_{g}/L_{s} $ where such description is not valid any more \cite{47} (here $ L_{s} $ is a length of 1D electrode of QPC, for the details of applicability of TLL description for actual experiments, see also Appendix A). Nevertheless, the "softened" versions of such steepest descent of $ Q(g) $ function are the ones, this paper is pretending to explain (such as ones on Fig.2a corresponded to temperatures of the order of $ 10^{1} mK$ which is the case in the experiments of Refs.[11,12]). In particular, since from Fig.2(a,b) it is evident that $ g_{cr} $ tends to be voltage-independent with the lowering of temperature, the value $ g_{cr} $ becomes a function of temperature only. Simple estimation of the first derivative $\partial Q/\partial g $ maximum from Eq.(17) in this low-temperature limit gives us a following expression for the "threshold" value $  g_{cr}$ of Luttinger liquid correlation parameter $ g $ as the function of other parameters in the system 
\begin{equation}
g_{cr}=g_{cr}(T) \approx \dfrac{\sqrt{3}\alpha}{2 \left\vert \dfrac{\Delta\lambda}{\Lambda} \right\vert \sqrt{\ln \left( \dfrac{\Lambda}{2\pi T}\right)}},   
\end{equation}
where $\vert \Delta \lambda \vert=\vert \lambda_{1} - \lambda_{2} \vert $ is the asymmetry in the electrostatic coupling of two charge states of charge-qubit to QPC detector, $ \Lambda \simeq \varepsilon_{F} $ is the high-energy cut-off of the order of Fermi energy in the 1D leads of QPC and $ 1/\sqrt{2}\leq \alpha \leq 1$ is a certain numerical fitting parameter obtained from the comparison of numerical value of $ g_{cr} $ on exact plots of Fig.2a,b with its analytical estimation of Eq.(19).
 
\begin{figure}
\includegraphics[height=13 cm,width=9 cm]{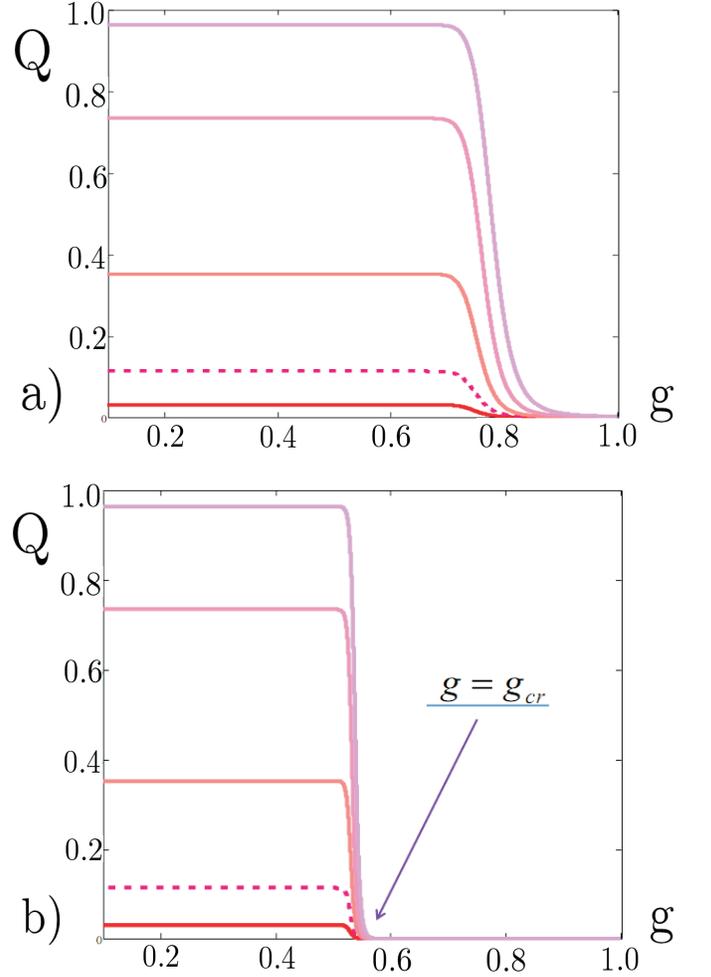}
\caption{\textit{Low-temperature quantum detector efficiency $ Q $ as the function of Luttinger liquid correlation parameter $ g $ according to exact Eq.(17) is plotted for different bias voltages in two temperature regimes. On Fig.2a  for all curves the temperature is equal to $ T=10^{-6}\Lambda $ (here $ \Lambda $ is the high-energy cut-off of the order of the Fermi energy in the leads of QPC) while the difference $ eV $between chemical potentials of QPC leads ($ V $ - is bias voltage) varies from 2 times smaller- (the lowest curve) to 8 times bigger (the top curve) than temperature $ T $ (i.e.  $ eV $ are equal to $ 0.5T;T;2T;4T;6T;8T $ from bottom to the top curve). On Fig.2b all curves are for the case where $ T=10^{-12}\Lambda $ (i.e. corresponding plots, in fact, model a zero-temperature limit of the theory) while voltages on Fig.2,b relate to the temperature in the same way as ones on Fig.2a. For both cases (Figs.2a,b) I put $ \Delta \lambda=0.2\Lambda $ and $ \Delta\tilde{\lambda}=0.02\Lambda $. Remarkably, on Fig.2b at $ T \rightarrow 0 $ one can observe a well-defined "threshold" value $ g_{cr} $ of Luttinger liquid correlation parameter which is approximately equal to $ 0.57 $ (for chosen parameters of the model). The abruptness of a "step" on the plot $ Q(g) $ around the "critical" value $ g_{cr} $ signals that a "quality" of quantum detection procedure is very unstable against small variations of local electron-electron interaction in QPC electrodes around respective "critical" magnitude  $ U_{cr} $ of local electron-electron interactions in QPC leads, corresponded to situation: $g \approx g_{cr} $.} 
}
\end{figure} 

\section{Discussion}

So, why at extremely low temperatures a quantum detection of charge-qubit state by means of QPC is near perfect when $ g<g_{cr} $ while at $ g>g_{cr} $ it fails according to Fig.2 ? - The answer is quite simple. There is a competition between two low-temperature processes where both them involve electron-electron interactions in the electrodes of QPC quantum detector, but in a different way.

First type of processes is evident from the structure of the formula (10). It represents Anderson orthogonality catastrophe \cite{51} between quantum states of charge-qubit and QPC quantum detector. Such type of decoherence can be described as a process of \textit{decoherence cloud formation} the latter "cloud" one may think of as of \textit{Kondo-like cloud of virtual QPC plasmons}\cite{55,56} in the vicinity of QPC tunnel contact (such clouds are schematically depicted on Figs.1a,b by means of a red-dashed ellipses in the QPC region). This polaronic Kondo-like cloud formation evidently takes exponentially large time(as well as it takes place in the case of true Kondo cloud in the Kondo-effect\cite{55,56}) $ \tau_{K}\simeq \tau_{ort,0}=\tilde{\Gamma}^{-1}_{K} $ - here $ \Gamma_{K} $ one can treat as a "width" of corresponding Kondo-resonance\cite{56}. But, remarkably, in the present model such the characteristic "Kondo" time - scales with Luttinger liquid correlation parameter $ g $ as $ \tau_{K}(g)\approx \left( \tau_{K}(1)\right)^{1/g^{3}} $, where $ \tau_{K}(1) $ is the corresponding characteristic time-scale for non-interacting fermions (i.e. one for QPC with Fermi-liquid leads, where $ g=1 $ ). The renormalization of corresponding contribution from Eq.(10) to the total decoherence rate by the parameter $ g<1 $ in the interacting case points out that a polaronic Kondo-like cloud formation (and, hence, the related decoherence) in the interacting case (i.e. when $ g<1 $) is much slower than in the case of non-interacting fermions (i.e. when $ g=1 $). The reason behind this behaviour becomes most clear in the special situation where $ g=1/n $, with integer $ n=1,2,3,.. $ associated with the case of charge-carriers fractionalization phenomena \cite{16} being exploited also in the case of Quantum-Hall edge states in the role of QPC electrodes. Indeed, from Eq.(10) and from all the above it follows that the total effective decoherence rate due to orthogonality catastrophe $ \tilde{\Gamma}_{K} $ is proportional to probability of Kondo-like resonance formation and can be written as $ \tilde{\Gamma}_{K}=\left( \tilde{\Gamma}_{K,p}\right)^{n}  $  for the case $ g=1/n $, where $ \tilde{\Gamma}_{K,p} $ is proportional to the probability of \textit{incomplete creation} of respective Kondo-like decoherence cloud for "interaction-dressed" electron on DQD. 

On the other hand, it is well-known\cite{53,55} that within the "bosonic language" the latter special "$ g=1/n $" case in the Luttinger liquid model represents a situation where a "bare" interacting electron being just a "kink" of bosonic field of $2\pi $ topological charge - is consisted of $ n $ statistically independent fractionally-charged bosonic quasiparticles (i.e. it consists of $ n $ fractionally-charged "kinks" of bosonic field, each of topological charge being equal to $ 2\pi/n $ ). Analogously, in the special case $ g=1/n $ ($ n=2,3,.. $) the above-mentioned polaronic Kondo-like cloud of plasmons, being an "overlap" between two thermal coherent bosonic states for two different moments of time, represents just \textit{a product of $ n $ identical fractionally charged polaronic clouds} which should emerge in order to provide a decoherence between "bare" electron of charge-qubit and the "bare" electron tunneling through the Luttinger liquid QPC. (Such plasmonic fractionally charged clouds for the case $ g=1/n $ (with $ n=3 $) were schematically shown on Fig.1,b by means of three red dashed ellipses.) Naturally, a simultaneous creation of $ n $ ($ n=2,3,.. $) Kondo-like "decoherence" clouds represents statistically much more rare event and, therefore, it takes much more time than a formation of just one such "cloud" in the $ g=n=1 $ case of Fermi liquid in QPC electrodes. Obviously, the same sort of qualitative arguments explains the effect from the orthogonality catastrophe contribution to decoherence in our system at arbitrary value $ g<1 $ of Luttinger liquid correlation parameter.
 
Second type of processes competing with the former in the expression (9) for quantum detector efficiency at low temperature is electron tunneling through the Luttinger liquid QPC. This tunneling is responsible for the acquisition of information about the charge-qubit quantum state and is captured in Eq.(9) by means of quantities $ W(\Delta\tilde{\lambda},T) $ and $\Gamma(\Delta\tilde{\lambda},T) $. As it follows from S-theorem and Eqs.(13-18), the exact low-temperature picture of the charge-qubit decoherence due to electron tunneling through Luttinger liquid QPC - totally coincides with Kane and Fisher classical prediction\cite{45,47} for the electron tunneling through the Luttinger liquid QPC in the "weak tunneling" regime. And this exact correspondence in the low-temperature limit represents another important consequence of S-theorem proven in this paper. Especially, Luttinger liquid effects of electron tunneling reveal themselves in the renormalization of tunneling rates on a very small (Kane-Fisher) factor $ \left(2\pi T/\Lambda_{g} \right)^{2/g-1}  $ which goes to zero when i) $ T \rightarrow 0 $ or ii) at strong electron-electron repulsion in the leads where $ g \rightarrow 0 $. This is simply a manifestation of well-known Kane-Fisher effect\cite{45,47}of strong suppression of electron density of states in the Luttinger liquid quantum wire with $ g<1 $ at Fermi energy (i.e. when $ T \ll \Lambda_{g} \simeq \varepsilon_{F}/g $ and $ T \rightarrow 0$). Obviously, from Eq.(18) it follows that electron tunneling through Luttinger liquid QPC with $ g<1 $ is also a very slow process. Especially, the characteristic time of electron tunneling $\tau_{t} $ scales with $ g $ as $ \left(\Lambda_{g}/2\pi T \right)^{2/g-1}  $ and, thus, extremely increases with the decrease of $ g $ (since $ g < 1 $). This behaviour perfectly agrees with the $ g $-dependence of characteristic time-scale $ \tau_{K} $ for the low-temperature orthogonality catastrophe discussed in the above (see Eq.(10)). (Notice also, that from Eqs.(17,18) it follows that in the special case where $ g=1/n $ the tunneling of a "bare" electron through a Luttinger liquid QPC proceeds also as a simultaneous tunneling of $ n $ statistically independent quasiparticles\cite{16} in according with  low-temperature picture of Kondo-like polaronic cloud formation in the case where $ g=1/n $ being discussed in the above of this Section.) 

The physical meaning of the sharp threshold between regions $ g<g_{cr} $ and $ g>g_{cr} $ in the behaviour of the function $ Q_{K}(g) $ near its critical value $ g=g_{cr} $ on Figs.2a,b as well as  physical meaning of its dependence (of Eq.(19) ) on other interaction parameters of the system and on the temperature (assuming that  temperature is small enough according to the limit $ T \rightarrow 0 $ ) - are both a subject of separate interest. In particular, such a "jump" in quantum detector efficiency $ Q $ near the value $ g_{cr} $ is a  fingerprint of certain electron-electron time-correlations collapse near the respective value $ U_{cr} \approx E_{F}\left( 1/g_{cr}^{2} -1 \right)$ of local electron-electron interaction in the QPC electrodes. Naturally, one may treat this collapse as the \textit{interaction-induced instability} of all the quantum detection procedure at extremely low temperatures when $ g \simeq g_{cr}$. Remarkably, this also means the instability of a "quantum limit" ( when $ Q \approx 1 $) of quantum detection for such strongly-interacting quantum systems.

To quantify somehow a competition between different quantum processes described here one need just to invert Eq.(19) in order to determine "critical" temperature $ T_{cr} $ of detection instability for any value of $ g $, i.e. one needs to know function $ T_{cr}(g) $. From Eq.(19) such estimation is straightforward
\begin{eqnarray}
T_{cr}=T_{cr}(g) \approx \dfrac{\Lambda}{2\pi}\exp \left[ -\frac{3\alpha^{2}}{4g^{2}}\left ( \frac{\Lambda}{\Delta\lambda}\right )^{2}\right ] \\
\approx 2 \left( \tilde{\Gamma}_{K}\right)^{g}= \tilde{\Gamma}_{K,p}. \nonumber  
\end{eqnarray}
 
In Eq.(20) $ \tilde{\Gamma}_{K} $ is the width of Kondo-resonance of Eq.(10) for charge-qubit quantum state being dressed into a Kondo-like cloud of virtual plasmons from QPC at given value of $ g $ , while $ \tilde{\Gamma}_{K,p} $ is a "partial" width of such resonance. Naturally,  $ \tilde{\Gamma}_{K} $ is proportional to the probability of "complicated" Kondo-cloud formation in the vicinity of QPC (such as a "cloud" on Fig.1,b consisted of three statistically independent polaronic clouds in the case of "charge-fractionalization" where $ g=1/3 $), whereas a partial width $ \tilde{\Gamma}_{K,p} $ - defines probability of \textit{incomplete} formation of such Kondo-like decoherence cloud , e.g. on Fig.1,b the latter case corresponds to situation where only one among three plasmonic clouds has emerged while the others are not yet(such situation is possible since one has $ \tilde{\Gamma}_{K,p} > \tilde{\Gamma}_{K}$ at $ g \neq 1 $ ). 

Thus, from Eqs.(19,20) one can conclude that situation $ g \approx g_{cr}(T) $ corresponds to condition $ T \approx T_{cr}\approx \tilde{\Gamma}_{K,p} $. And since all the graphs on Fig.2a(b) are plotted for single fixed value of temperature $ T $ (being equal to $ 10^{-6}\Lambda $ for Fig.2,a and to $ 10^{-12}\Lambda $ for Fig.2,b ), it becomes clear that parameter region $ g < g_{cr}(T) $ on Figs.2a,b corresponds to the case $ \tilde{\Gamma}_{K} < \tilde{\Gamma}_{K,p} < T $, whereas the region $ g_{cr}(T) < g \leq 1$ on both Figures 2a,b  -describes a "Kondo-resonance" situation, where $ T < \tilde{\Gamma}_{K,p}$. 

Now, recalling that as it follows from Eqs.(7,10) our analysis is valid only for time intervals $ t \leqslant \tau_{meas} $ restricted by the condition $ \tau_{meas} \leqslant \hbar/T $ ($ T\rightarrow 0 $) we can "translate" all possible situations described by Figs.2a,b on the "language" of timescales of corresponded physical processes. For example, one can see that parameter region $ g < g_{cr}(T) $ on Figs.2a,b corresponds to the case where $ \tau_{meas} < \tau_{K,p} < \tau_{K}  $ i.e. when the maximal measurement time $ \tau_{meas}  $ is still smaller than a characteristic time $  \tau_{K,p} $ of only \textit{partial} emergence of Kondo-like decoherence cloud. On such the short time-scales no one Kondo-like cloud of virtual plasmons has enough time to "condense" around the QPC region and, hence, all decoherence in the system is due to electron tunneling through QPC which is also an issue of information about charge-qubit quantum state. That is why in this parameter region on Figs.2a,b the detector efficiency $ Q $  is near perfect (for high enough bias voltages). On the other hand, for the rest of entire interval of $ g $ values, when $ g_{cr}(T) < g \leq 1$ on both Figures 2a,b  one deals with situation, where either $  \tau_{K,p} < \tau_{K} < \tau_{meas}$ or $  \tau_{K,p} \leqslant \tau_{meas} < \tau_{K}$. Here the first case refers to physical situation similar to one of Kondo-resonance, i.e. when the measurement time $ \tau_{meas} $ is longer than the characteristic time $ \tau_{K} $ of formation of Kondo-like decoherence cloud(s) (those are depicted on Fig.1 by means of red dashed ellipses), hence, such clouds have enough time to emerge and to decohere charge-qubit quantum state due to Anderson orthogonality catastrophe irrespective of any electron tunneling processes in the system. This results in very fast decoherence of charge-qubit as compared to acquisition of information about its state, and leads to a breakdown of all the quantum detection procedure at $ g_{cr}(T) < g \leq 1$. 

But the most remarkable regime is when $  \tau_{K,p} \lesssim \tau_{meas} < \tau_{K}$  which corresponds to a narrow region around critical value of correlation parameter on Figs.2a,b, where $ g \simeq g_{cr}(T) $ (or equally $ T \simeq T_{cr}(g) $). The latter case describes an "intermediate" time-scale of measurement where complicated (i.e. like one on Fig.1,b) Kondo-like decoherence cloud has  formed only \textit{partially} (e.g. when only one among three virtual plasmonic clouds on Fig.1,b has the time to emerge while the other "clouds" do not). Obviously, on such intermediate time-scales the probability of the existence of \textit{complete} Kondo-like decoherence cloud   and the probability of its absence - both are non-zero. Hence, corresponding Kondo-like cloud formation time-scale $ \tau_{K} $ becomes uncertain when $  \tau_{K,p} \lesssim \tau_{meas} < \tau_{K}$ which makes uncertain also the quantum detector efficiency $ Q $ in the vicinity of critical interaction $ U_{cr} $  as it can be seen from Figs.2a,b at $ g \simeq g_{cr}$. This fact, in turn, leads to the interaction-induced instability for all the procedure of quantum detection (including its quantum limit) in a narrow parameter "window" around the critical value $ g_{cr}(T) $ of Luttinger liquid correlation parameter in the QPC leads. 

Now we are at the point where too large decoherence time-scales measured in two experiments on charge-qubit manipulation of Refs.[11,12] can be explained within the theory developed in this paper. This is because as I have demonstrated here, it is very plausible that authors of Refs.[11,12] had measured not "true" decoherence timescales  $ \tau_{dec} $ of their charge-qubits but only corresponding acquisition of information time-scales $ \tau_{acq} $ (just by construction of those two experiments). And as it was already mentioned in the above (and in the Appendix A), the main source of decoherence in the case of both these low-temperature experiments should be referred to interaction between charge-qubit and QPC (or SET) quantum detector. Therefore, the ratios between the initially expected and measured decoherence time-scales in the experiments of Refs.[11,12] should be equal to ratios between $ \tau_{dec} $ and $ \tau_{acq} $ i.e. to \textit{quantum detector efficiences} $ Q $ in those experiments. Hence, authors of Ref.[11] have reported that $ \tau_{dec}/\tau_{acq}=Q \approx 10^{-2} $ whereas for a bit different experiment of Ref.[12] it has been claimed that $ \tau_{dec}/\tau_{acq}=Q \approx 10^{-1} $. Since the temperature regime of both experiments of Refs.[11,12] agrees with one on Fig.2,a ($ \simeq 10^{1}- 10^{2}mK $), one could check if this figure includes measured values ($ 10^{-2} $ and $ 10^{-1} $) of the efficiency $ Q $  of QPC quantum detectors, those being realized in the experiments of Refs.[11,12].  Indeed, as one can see from Fig.2,a these values of quantum detector efficiency correspond to the value $ \simeq 0.9 $ of Luttinger liquid correlation parameter $ g $ in QPC electrodes for experiment of Ref.[11] and to the value $ \simeq 0.85 $ of this parameter in QPC leads for the experiment of Ref.[12]. Remarkably, both these values (as it can be seen from Fig.2,a) are very close to "critical" value $ g_{cr} $ of this correlation parameter within given temperature regime. 

In other words, in the framework of the theory developed in the above a huge difference between  observed and expected decoherence time-scales in the experiments of Refs.[11,12] can be referred completely to the \textit{instability of all the quantum detection procedure} by means of QPC quantum detectors. In more details, in both cases\cite{11,12} experimentalists could use for their QPC quantum detectors the electrodes which should be treated  as quasi-one-dimensional quantum wires with the properties of a weak electron-electron repulsion therein (e.g. when $ g \simeq 0.9$) rather than simple "bulk" Fermi-liquid metallic leads with $ g=1 $. This is very plausible just because the exact value of Luttinger liquid correlation parameter $ g $ for given ballistic quantum wire is unknown \textit{a-priori} and can be clarified more-less precisely only in the process of measurement of ballistic current through this wire. On the other hand, the good enough agreement between temperature regimes of the experiments of Refs.[11,12] and temperature regime of Fig.2,a -provides that measured values of $ Q $ and respective values of $ g $ taken from Fig.2,a  - correspond to the narrow region of quantum detection instability around  certain critical value $ g_{cr} $ from Fig.2,a. Evidently, in such the instability region small variances in values of Luttinger liquid correlation parameter $ g $ in QPC electrodes lead to huge changes in the efficiency $ Q $ of respective QPC quantum detector and result in huge difference between decoherence- and acquisition of information time-scales like in the experimental situations of Refs.[11,12]. This key observation seems to be very important practical outcome from all the theory being developed in the above. 

To conclude, in this paper the properties of low-temperature quantum detection of charge-qubit quantum state by means of Luttinger liquid QPC were considered. Since in zero-temperature limit both charge-qubit decoherence rate and the acquisition of information rate for Luttinger liquid QPC tend to zero in the lowest order in electrostatic interaction between two, the corresponding decoherence- and acquisition of information timescales both diverge leading to uncertainty of QPC detector properties within this approach. However, in the above it was shown (by means of S-theorem and S-lemma) that corresponding Keldysh-contour-ordered T-exponent for Luttinger liquid QPC can be calculated exactly, providing its exact re-exponentiating to known result. These rigorous mathematical statements being useful for a number of problems involving Luttinger liquid non-equilibrium dynamics also provide us with exact expressions for charge-qubit decoherence rate  and acquisition of information rate. In turn, both these exact quantities reveal exact picture of quantum detection by means of Luttinger liquid tunnel junction in the low- (and zero-)temperature limit. Especially, with the help of latter exact results, it was established that in the low-temperature limit there exists definite temperature-dependent critical value $ g_{cr} $ of Luttinger liquid correlation parameter, which separates region of near-perfect quantum detection at $ g < g_{cr} $ from the region of quantum detection breakdown at  $ g_{cr} < g \leq 1$. Moreover, the obtained exact results for low-temperature detection by means of Luttinger liquid QPC reveal the existence of \textit{instability} in quantum detector properties for any Luttinger liquid QPC at $ g \approx  g_{cr}(T) $ in the low-temperature limit, including the possibilty for the system to have the \textit{unstable} quantum limit of detection. As well, I discussed here the mapping of theoretical model under investigation on two existing experiments with charge-qubit dynamics. Surprisingly, in this context it was found that unclear  mismatch between expected and measured decoherence timescales in both these experiments could be successfully explained in the framework of the low-temperature "detection instability" effect which has been revealed for the first time in this paper. Thus, a paper contains novel quantum mesoscopic effects which, on one hand, are of fundamental theoretical importance and, on the other hand, are experimentally robust regarding implementations to any mesoscopic devices, those involving  quantum-point contacts with Luttinger- (and Fermi-)liquid leads. Especially, all the described physics can be realized in future modifications of the actual experiments on quantum detection and decoherence control at temperatures close to absolute zero. Interaction effects in quantum detection being established here should play a significant role in a wide range of physical problems concerning quantum feedback and control in quantum nanoelectronics. 
\section{Acknowledgements}
Author would like to thank Yuval Gefen, Dimitry Gangardt and Daniel Loss for discussions on a broader context of the obtained results.

\appendix

\section{Relation of underlying theoretical model to the actual experimental background}\label{sec:just}

It is widely known (see e.g. a classic work by Kane and Fisher\cite{47}) that an inevitable  constriction of QPC electrodes in the vicinity of QPC tunnel contact in many cases implies an effective one-dimensional (1D) geometry of QPC electrodes which, in turn, might result in the appearance of strong electron repulsion in such 1D leads of QPC detector\cite{47}. (On Fig.1 corresponded constriction of QPC electrodes is shown by means of two narrow cylindrical wires. Notice, that in the experimentally possible case of metallic carbon SWNT electrodes the correspondence between Fig.1 and real device geometry becomes explicit.) In such the case a local repulsive electron-electron interaction becomes sufficient for 1D electrodes with relatively low concentration of charge carriers\cite{47} such as, for example, some types of single-walled carbon nanotubes (SWNTs) and GaAs nanowires\cite{17,42,43,44} (e.g. GaAs compound was used for QPC electrodes in the experiment of Ref.[12]). It is widely known, that 1D systems of both non-interacting and interacting fermions are perfectly described by the model of Tomonaga-Luttinger liquid (TLL)\cite{45,46,47,53,54,55} if temperatures and all the other characteristic energies in the system are much smaller than $ E_{F} $ - a Fermi energy of a given quantum wire (hence, $ E_{F} $ - renormalized by the interaction parameter $ g $ - usually serves as the high-energy cut off $ \Lambda_{g}=E_{F}/g $ for the linearized energy spectrum of the theory and typically does not exceed the value of several electron-volts). Here $ g $  - is the dimensionless TLL  correlation parameter which relates to the characteristic potential energy of local electron-electron repulsion $ U_{s} $ in a given 1D quantum wire as $ g \approx \left( 1 + U_{s}/2E_{F} \right) ^{-\frac{1}{2}}$ . Parameter $ g $ is confined in the limits: $ 0 < g \leq 1 $ and serves as the only measure of electron-electron interaction in the bosonized version of TLL theory; it renormalizes all group velocities, cutoff energies and defines  critical exponents for all Luttinger liquid correlators\cite{47,53}. Notice, that value: $ g=1 $ corresponds to non-interacting fermions, describing the case where the electrodes of QPC can be represented within usual Fermi liquid (FL) picture. At the same time the limit $ g \rightarrow 0$ describes opposite situation of infinitely strong local electron repulsion in the one-dimensional QPC leads. That is why the concrete value of $ g $ in some sense defines an "effective dimensionality" of QPC electrodes: in the case $ g=1 $ one may treat them as usual 3D bulk metal while in the case $ g <1 $ - the one-dimensional model for QPC leads is demanded (the former case is depicted  on Fig.1,a while the latter - on Fig.1,b).  

In most experimental realizations of quantum-point contacts including ones from Refs.[11,12] the characteristic lengths $ L_{s} $ of respective QPC electrodes can vary from approximately $ 1 \mu m $ to $ 10 \mu m $, while their widths vary from $ 1 nm $ to $ 10nm $. At the same time the lattice constant $ a_{0} $ in all quantum wires varies (approximately) between $ 0.4 nm$ and $0.7 nm$. The ratio between these two length-scales defines a relation between the upper energy bound of the Luttinger liquid description ( this bound is a high-energy cut off $ \Lambda_{g} $) and the lower energy bound (which is the energy level spacing due to finite lengths of quantum wire $ \varepsilon_{min}$). This ratio reads $ a_{0}/L_{s}=\varepsilon_{min}/\Lambda_{g} \ll 1 $. For example, in the experiments of the interest\cite{11,12} minimal possible value for this ratio is of the order of $ 10^{-5} $ which is of the same order as the ratio: $ T/\Lambda_{g} $ in these experiments. However, it is known that there exists one more restriction on the applicability of the TLL model to the concrete quantum wire\cite{47}. The TLL model is valid at low temperatures which though should not be smaller than the energy level spacing due to finite lengths of the QPC electrodes\cite{47}(otherwise thermal fluctuations in the wire are unable to provide a continuous spectrum needed for TLL model) $ T > \varepsilon_{min}=\hbar v_{g}/L_{s} $. (here $ v_{g}=v_{F}/g $ - is the group velocity of longitudinal plasmonic excitation in quantum wire; it represents just a Fermi velocity renormalized by the electron-electron interaction).  Naturally, from the above one can conclude that  all the characteristic energy scales of a given QPC should fulfil following basic inequality: $ \varepsilon_{min} < T, eV \ll \Lambda_{g}$ to be described within TLL model. Here $ V_{g,i}=1/e\left( e \varphi_{i} - E_{F}\right) =\varphi_{i} $ is a gate voltage on the $ i $-th gate on Figs.1a,b ($ \varphi_{i} $ is electrostatic potential of corresponded gate on Figs.1a,b , $ i=1,2,3 $ - is index numbering gates on Fig.1).   It seems, that for QPC quantum detectors from experiments of Refs.[11,12] the latter inequality is satisfied. Thus, everywhere in my model I will presume the validity of the latter basic inequality which allows for the description of QPC quantum detector in terms of Luttinger liquid tunnel junction\cite{47}. 

The quantum-point contact under consideration is biased. It means that its "left" and "right" 1D electrodes are coupled from their "free" ends with "left" and "right" bulk reservoirs of non-interacting electrons each - with definite constant electrochemical potential $ \mu_{L(R)} $ (the constant difference between them is bias- (or driven) voltage: $ V=(\mu_{L} - \mu_{R})/e $, $ eV \ll \Lambda_{g} $), those are depicted as "source" and "drain" on Fig.1a,b. Thus, tunnel current constantly flows through a tunnel junction creating, in general, a non-equilibrium dynamics of interacting electrons in both QPC electrodes. The related time-dependent non-equilibrium effects are the subject of intense theoretical studies by themselves\cite{7,8,16,17}.  However, below we can safely neglect all these time-dependent nonequilibrium effects in the leads assuming that all electrons in both QPC electrodes have the same constant electrochemical potentials $ \mu_{L(R)} $ as ones of respective bulk electron reservoirs ("source" and "drain" on Fig.1). This simplification is well-grounded due to the fact that because of a very large number of electrons in QPC electrodes, adding one extra low-energy electron to the lead (or removal electron from it) in the regime of ballistic electron transport through the QPC (this is the case e.g. in the experiments of Refs.[11,12]) - disturbs  many-electron quantum state of a given wire only a little, resulting only in the negligibly small disturbance in the equilibrium values of respective electrochemical potentials of the leads. One may expect that corresponding non-equilibrium corrections are of the order of negligible effects associated with finite lengths of given 1D electrodes of QPC. So, this assumption is valid as long as we have: $ eV,T,eV_{g,i} > \hbar v_{g}/L_{s} $ (where $ L_{s} $ - is the length of QPC electrode).i.e. this assumption is compatible with the validity of Luttinger liquid model for QPC leads\cite{47,50}. 

Now let us consider the Hamiltonian of charge-qubit $H_{DQD}$ and Hamiltonian  $H_{int}$ of electrostatic interaction between charge-qubit and QPC quantum detector. In the experiments of Refs.[11,12] charge-qubit represents a quantum state of one excess electron "shared" between two lowest-unoccupied energy levels of two overlapping quantum dots which both form a double-quantum-dot (DQD) system. The double-quantum-dot (DQD) in both experimental situations represents an artificial "molecule"- an isolated Phosphorous-doped silicon-\cite{11} or GaAs- \cite{12} "droplet" whose excess electrons are confined within effective double-well electrostatic potential influenced by several external gates (one can see details on the electron micrograph pictures in Refs.[11,12]).

To control the total number of electrons on DQD "droplet" one can use either a famous Coulomb blockade phenomenon\cite{18} or just a large spacing in the levels of spatial quantization in a given DQD. Especially, in all the experiments with charge-qubits one should have : $ T, eV \ll max\left[E_{c,n}, \Delta_{n} \right] $ (I presume the validity of this condition throughout in this model), where $ E_{c,n} \approx N_{n}e^{2}/D_{qd,n} $ -is the Coulomb charging energy of DQD droplet (index $ n=1,2 $ numbers quantum dots in given DQD structure, $ eN_{n} $ is a total electric charge of $ n $-th quantum dot and $ D_{qd,n} $ is a characteristic diameter of $ n $-th quantum dot in given DQD structure), while $ \Delta_{n} \approx \hbar v_{F}/D_{qd,n} $ is the electron level spacing in each quantum dot of DQD structure due to spatial quantization. For the nanoscale-sized DQD droplets, in the absence of external magnetic fields - which is the case for both experiments of Refs.[11,12], one usually has the situation where $  \Delta_{n} \ll E_{c,n}$ provided that charging energy of DQD is of the order of the Fermi energy of QDC electrodes: $ E_{c,n} \sim E_{F} $ which is usually of the order of several electron-volts\cite{11,12} and, thus, a Coulomb blockade remains a main tool to control electron number in DQD.

The presence of the three capacitive gates in our model (as well as in the experiments of Refs.[11,12]) means that in our case: $ \varepsilon_{1}=\varepsilon(V_{g,1},V_{g,2}) $ and  $ \varepsilon_{2}=\varepsilon(V_{g,2},V_{g,3}) $, while for inter-dot tunnel coupling one has in general $ \gamma=\gamma(V_{g,1},V_{g,2},V_{g,3}) $ (see Fig.1). Here $ V_{g,i}=1/e\left( e \varphi_{i} - E_{F}\right) =\varphi_{i} $ is the gate voltage on the $ i $-th gate on Figs.1a,b ($ \varphi_{i} $ is the electrostatic potential of corresponded gate on Figs.1a,b , $ i=1,2,3 $ - is the index numbering gates on Fig.1). Hence for all gate voltages of the model one should have $ eV_{g,i} \lesssim E_{c,n} $ provided that in typical experiments like ones from Refs.[11,12] $ eV_{g,i} $ can vary from $ 0.1 eV$, see Ref.[12]  to $ 10^{1} eV $, see Ref.[11].

The described above charge-qubit control scheme have been realized in the experiments of Refs.[11,12] would result in the controlled time-dependent "floating" of one excess electron quantum state within the DQD "droplet" area in the absence of its interaction with QPC quantum detector. However, this "floating" of DQD excess electron quantum state is precisely the only key to readout the time-dependent quantum state of corresponding charge-qubit in the procedure of quantum measurement\cite{33,34,35,36}. For this purpose one needs only to couple  the current-carrying QPC  capacitively to the DQD system and then an effective tunnel barrier of QPC will "feel" (only statistically, of course, as we have here purely quantum-mechanical description of electron dynamics) any spatial deflections of the electric charge within the DQD area. 

In such a way, measuring repeatedly charge transfer through the QPC which electrostatically interacts with charge-qubit, the authors of Refs.[11,12] indirectly read out\cite{48,49} the information about the charge-qubit quantum state at any instant of time between $ t=0 $ and $ t'=\Delta t_{p}  $ (For this purpose they apply to DQD one more "control" gate which produces a "train" of voltage pulses of definite duration $ \Delta t_{p} $ , hence, in order to plot the respective graphs experimentalists should only vary the duration $ \Delta t_{p} $ of each pulse. (However, the amplitudes of latter pulses are small comparing with another gate voltages; these amplitudes vary from $ 10^{2} mV $ in Ref.[11] to $ 10^{2} \mu V  $ in Ref.[12]).) Resulting plots are ones for measured occupation probabilities of two base charge-qubit quantum states ($\vert 1 \rangle  $ and $ \vert 2 \rangle $ - each state corresponds to one energy level in a two-level quantum system) as the functions of time (or to be precise, as the functions of $ \Delta t_{p} $ ). Corresponding graphs are the most remarkable experimental achievements of Refs.[11,12]. Generally speaking, most experimental graphs of Refs.[11,12] represent just a picture of Rabi oscillations of electron quantum state in the  two-level quantum system of DQD (i.e. of charge-qubit), but with the amplitudes  \textit{decreasing with time} more or less intensively\cite{11,12}. Hence, the measured damping of Rabi oscillations in both experiments \cite{11,12} is a clear fingerprint of an inevitable and uncotrollable decoherence\cite{37,38,39,40,41} which assists any process of quantum measurement due to necessary interaction between quantum measured system (electron on the DQD) and quantum detector (electrons tunneling through quantum-point-contact)\cite{40,41,50}.

The last term $ H_{int} $ in the total Hamiltonian $ H_{\Sigma} $ describes electrostatic (i.e. density-density) repulsive interaction between the charge-qubit on DQD and the electron, which tunnels through QPC quantum detector\cite{50}. This corresponds to the actual experimental situations of Refs.[11,12]. According to our previous work of Ref.[50] here in this paper it is assumed the locality of this interaction. One may argue that in the reality of experiments [11,12] such interaction is generally non-local (i.e. it involves a certain area of QPC electrodes in the vicinity of their tunnel coupling ). However, a typical tunneling length for the most QPC is of the order of lattice constant $ a_{0} $ of its electrodes and in the above it was explained that one can safely neglect the finiteness of this parameter taking the limit $ a_{0} \rightarrow 0 $ of the validity of continuous TLL description for QPC electrodes\cite{47,53}. In this limit QPC tunnel junction is well-described\cite{47}  by means of only local values $ \theta_{L,R}(x=0,t) $ and $ \varphi_{L,R}(x=0,t) $ of corresponding bosonic fields on the edges of QPC electrodes at one point $ x=0 $ (see Refs.[47,50]). Within this description all the "non-locality" of the electrostatic interaction between $ n $-th quantum dot of DQD ($ n=1,2 $) and QPC tunnel junction is encoded in the concrete numerical value of corresponding "bare" tunneling constant $ \tilde{\lambda}_{n} $. Notice also, that the theoretical model of Luttinger liquid QPC in this paper coincides with one from classical Kane and Fisher paper in the case of "weak" tunneling through TLL tunnel junction\cite{47} the same approach was also used in our paper\cite{50} for finite temperature analysis of this model. 

Both coupling constants $ \lambda_{n} $ and  $ \tilde{\lambda}_{n} $ describe the same electrostatic interaction, however, due to "weak tunneling" nature of coupling constants $ \tilde{\lambda}_n $ the latter should be much smaller than respective constants $ \lambda_n $ of purely electrostatic density-density interaction: $ \tilde{\lambda}_n \ll \lambda_n $.  Remarkably, for the DQD subsystems deposited in the close vicinity of their QPC quantum detector (as it takes place in both experiments of Refs.[11,12]), the magnitudes of both coupling constants related to the same $ n $-th quantum dot should be comparable with the Coulomb charging energy $ E_{c,n} $ of this quantum dot: $ \tilde{\lambda}_n , \lambda_n \lesssim E_{c,n}$. It means that underlying model includes the case where given charge-qubit and its quantum detector can be, indeed, strongly interacting with each other.  As to the  symmetry between the QPC couplings  to 1-st and to 2-nd quantum dots, one could claim, that under the "resonance" condition of our model: $ \varepsilon_{1}=\varepsilon_{2}=0 $ and $ \gamma \rightarrow 0 $) such the asymmetry should be quite weak, i.e. $ \tilde{\lambda}_1 \simeq \tilde{\lambda}_2 $ and $\lambda_1 \simeq \lambda_2  $. This situation, of course, complicates the discrimination between two quantum states: $ \vert 1 \rangle $ and $ \vert 2 \rangle $ in the process of quantum detection \cite{50}. However, a more favourable experimental situation where $ \tilde{\lambda}_{1(2)} < \tilde{\lambda}_{2(1)}  $ and $ \lambda_{1(2)} < \lambda_{2(1)}  $  is possible if one manages to deposite a DQD system in such a way, that one quantum dot of DQD would be closer to QPC than another one (see e.g. Fig.1 and Refs.[11,12]). The model under consideration authomatically includes both "symmetric" and "asymmetric" possibilities (see e.g. Eq.(4)).

\section{The proof of S-Theorem and S-Lemma}\label{sec:Reexp_tunn}

$ \triangledown $ \textit{The proof of S-Theorem.}
First, expanding only time- and anti-time-ordered parts in Keldysh-contour-ordered exponents in Eq.(8) (for the rest of the contour, see the end of the proof of S-lemma below) one has
\begin{align}%\label{eq:expd}
\begin{split}
\tilde{Z}_{12(21)}(t)= \langle \mathcal{T}_K \lbrace  \sum_{n=0}^{\infty} \int^t_0 d \tau_{1}  \int^{\tau_{1}}_{0} d \tau_{2} \ldots \int^{\tau_{n-1}}_{0} d \tau_{n} \\
(i\tilde{\lambda}_{1(2)})^{n} A_{0}(\tau_{n})\ldots A_{0}(\tau_{1})\rbrace  \times \\
\lbrace  \sum_{n'=0}^{\infty} \int^t_0 d \tau_{1}  \int^{\tau_{1}}_{0} d \tau_{2} \ldots \int^{\tau_{n'-1}}_{0} d \tau_{n'} \\
(-i\tilde{\lambda}_{2(1)})^{n'} A_{0}(\tau_{1})\ldots A_{0}(\tau_{n'})\rbrace \rangle .
\end{split}
\end{align}
with $ A_{0}(\tau)=\cos{\left[\varphi_-(\tau)+eV\tau\right]} $. Now to apply properly Wick theorem to the latter expansion, let us consider its "building block" - the two-time correlator $ \langle  A_{0}(\tau_{n}) A_{0}(\tau_{n'})\rangle  $ with $ A_{0}(\tau)=A_{0}(\tau)=\cos{\left[\varphi_-(\tau)+eV\tau\right]}$. Taking into account the facts that $ \langle\sin{\left[\varphi_-(\tau)\right]}\rangle=0 $ because $ \langle\varphi_-(\tau)\rangle=0 $ by definition of fluctuating $ \varphi_{-} $ quantum field and $ \langle\cos{\left[\varphi_-(\tau)\right]}\rangle=0 $ because $\langle e^{\pm i\varphi_-(\tau)}\rangle=0 $ and that only "neutral" configurations of the kind $ \langle e^{i\varphi_-(\tau)}e^{-i\varphi_-(\tau')}\rangle=\langle e^{-i\varphi_-(\tau)}e^{i\varphi_-(\tau')}\rangle $ do not vanish (see e.g. Ref.[53]). All this allows us to write
\begin{align}%\label{eq:corr}
\langle A_{0}(\tau_{n}) A_{0}(\tau_{n'})\rangle=f_{g}(\tau_{n},\tau_{n'}) \cos[eV(\tau_{n}-\tau_{n'})],
\end{align} 
where, using well-known Baker-Hausdorf formula for the averages of two operator-valued exponents,  
\begin{align}
\begin{split}
 f_{g}(\tau_{n},\tau_{n'})=\langle e^{i\varphi_-(\tau)}e^{-i\varphi_-(\tau')}\rangle \\
 =e^{\frac{-\left\langle\left[\varphi_{-}(\tau_{n})-\varphi_{-}(\tau_{n'})\right]^2\right\rangle}{2}}e^{\frac{\left[\varphi_{-}(\tau_{n}),\varphi_{-}(\tau_{n'})\right]}{2}} .
\end{split}
\end{align}
For the commutator in the latter equation using standard Fourier decomposition of bosonic phase-fields\cite{53} one has 
\begin{align}
\begin{split}
\left[\varphi_{-}(\tau_{n}),\varphi_{-}(\tau_{n'})\right]=-2i\vartheta_{g}\sgn{\left[\tau_{n}-\tau_{n'}\right]}.
\end{split}
 \end{align}
where $ \vartheta_{g}=const $ (notice, that in our particular case $ \vartheta_{g}=\pi/g $ - see e.g. Refs.[50,53]).  Taking into account this expression, it is possible to rewrite our basic pair correlator in the following "symmetrized" form which is more suitable for our calculations
\begin{align}%\label{eq:defav}
\langle  A_{0}(\tau_{n}) A_{0}(\tau_{n'})\rangle = \langle  A_{0}(\tau_{n}) A_{0}(\tau_{n'})\rangle_{S} \cdot 
\left\{ 
 \begin{matrix}
    e^{-i\vartheta_{g}} &, \tau_{n} > \tau_{n'} \\
    e^{i\vartheta_{g}}  &,  \tau_{n} < \tau_{n'} \\
    1 &, \tau_{n} = \tau_{n'}
    \end{matrix} \right.
\end{align}
In above equation I defined following "symmetrized" two-time correlator 
\begin{align}%\label{eq:symm}
\begin{split}
\langle  A_{0}(\tau_{n}) A_{0}(\tau_{n'})\rangle_{S}=\langle  A_{0}(\tau_{n'}) A_{0}(\tau_{n})\rangle_{S} \\
=u(\tau_{n}-\tau_{n'})\cos{\left(f(\tau_{n})-f(\tau_{n'})\right)}
\end{split}
\end{align}
which is the even function of the time difference $s=(\tau_{n}-\tau_{n'}) $. Even function $ u(s) $ in turn is nothing more than average
\begin{align}%\label{eq:uf}
\begin{split}
u(\tau_{n}-\tau_{n'})=e^{\frac{-\left\langle \left[\varphi_{-}(\tau_{n})-\varphi_{-}(\tau_{n'}) \right]^2\right\rangle}{2}}=e^{\frac{-I(\tau_{n}-\tau_{n'})}{2g}} .
\end{split}
\end{align}
For the integral from bosonic average in the exponent of Eq.(B7) one can obtain\cite{50,53} 
\begin{align}
\begin{split}
\left\langle\left[\varphi_-(\tau_{n})-\varphi_-(\tau_{n'})\right]^2\right\rangle =\frac{I(\tau_{n}-\tau_{n'})}{g} =\\
 \frac{4}{g}\mathcal{P}
  \int^{\infty}_0 \frac{ \rm{d}k}{\left| k \right|} e^{- \alpha_0 \left| k
  \right| }\left[2n_b(k)+1\right] \left[ 1 - \cos \left( v_g  \left| k \right|(\tau_{n}-\tau_{n'}) \right) \right] .
\end{split}
\end{align}
%This allows us to write $f(\tau_{n},\tau_{n'})=f(\tau_{n}-\tau_{n'})$ .
Making use of Eqs.(B2-B5) together with our basic expansion (B1) for $ \tilde{Z}_{12(21)}(t) $ one can rewrite Eq.(8) as a power series over even natural $ n $ from zero to infinity, where each term of the $ n $-th order in the infinite sum consists of all possible combinations of  $ n $-point correlators ( $ n $- even natural number) being constructed from  $ n $ operators $ A_{0}(\tau_{k}) $ ($ k=1,..,n $). As the result, one obtains following exact expansion
\begin{align}%\label{eq:comb}
\begin{split}
\tilde{Z}_{12(21)}(t)=1+\sum_{n=2}^{even}  
\sum_{j,k=0}^{n}C^{n}_{k} 
 e^{i \vartheta_{g} k}e^{-i \vartheta_{g}(n/2-k)} \\
    \times C^{n}_{j}
   (i\tilde{\lambda}_{1(2)})^{(j+k)} 
    (-i\tilde{\lambda}_{2(1)})^{(n-(j+k))} \\
 \times \int^{t}_{0} d \tau_{1}\int^{\tau_{1}}_{0} d \tau_{2}\ldots \int^{\tau_{k-2}}_{0} d \tau_{k-1}\int^{\tau_{k-1}}_{0} d \tau_{k} \\
 \times \int^{t}_{0} d \tau'_{1}\int^{\tau'_{1}}_{0} d \tau'_{2}\ldots \int^{\tau'_{(n-k)-2}}_{0} d \tau'_{n-k-1} \int^{\tau'_{(n-k)-1}}_{0} d \tau'_{n-k}\\
\times \langle \mathcal{T}_K A_{0}(\tau_{k})\ldots A_{0}(\tau_{1})A_{0}(\tau'_{n-k})\ldots A_{0}(\tau'_{1})\rangle_{S} . 
\end{split}
\end{align}
Here $C^{n}_{k}=\frac{n!}{k!(n-k)!}$ are the standard binomial coefficients being equal to the number of ways one could select $ k $ time-ordered operators $A_{0}(\tau_{1})\ldots A_{0}(\tau_{k}) $ from $ n $  operators $A_{0}(\tau_{1})\ldots A_{0}(\tau_{n}) $ (then the product of the remained $ n-k $ operators $ A_{0}(\tau_{1})\ldots A_{0}(\tau_{k}) $ in each $ n $-th order will be automatically anti-time ordered after such the selection, since in the formulas (B1-B9) we do not consider terms with non-ordered operators - the latter contributions will be considered below ), while $C^{n}_{j}$ counts the number of situations  where $ \tau=\tau' $ in each $ n $ -order term. Also $\left\langle A_{0}(\tau_{m})\ldots A_{0}(\tau'_{n})\right\rangle_{S}$ in Eq.(B9) represents the "symmetric" part of the correlator $\left\langle A_{0}(\tau_{m})\ldots A_{0}(\tau'_{n})\right\rangle$ with respect to exchange $\tau_{m} \leftrightarrow \tau'_{n} $ (or $\tau_{m} \leftrightarrow \tau_{n} $, or $\tau'_{m} \leftrightarrow \tau'_{n} $) for any pair of time variables $\tau_{m}$ and $\tau'_{n}$ (or $\tau_{m}$ and $\tau_{n}$, or $\tau'_{m}$ and $\tau'_{n}$) from Eq.(B1). The latter property is because all the effect from these exchanges is already taken into account by means of phase factors in Eq.(B5,B9) due to the time- and anti-time ordering in Eq.(B1), thus, such  permutation of variables in Eq.(B9) should not change the result of integrations in Eq.(B1). 

To proceed further, let us prove the S-lemma mentioned in the main text.

$ \triangledown $ \textit{The proof of S-Lemma.} Obviously, one can write the most general result of the averaging $\left\langle A_{0}(\tau_{k})\ldots A_{0}(\tau_{1})\right\rangle_{S}$ using only two properties: i)the symmetry of the correlator: $\left\langle A_{0}(\tau_{k})\ldots A_{0}(\tau_{1})\right\rangle_{S}$ with respect to exchange $\tau_{m} \leftrightarrow \tau_{n} $ for any $ n $ and $ m $; ii) the fact that only "neutral" combinations  $ (\varphi_{-}(\tau_{k})- \varphi_{-}(\tau_{k-1})+\ldots+\varphi_{-}(\tau_{k})- \varphi_{-}(\tau_{1}))$  of bosonic field $\varphi_{-}(\tau_{k})$ with the same field at another moments of time $\varphi_{-}(\tau_{k-1})\ldots\varphi_{-}(\tau_{1})$  do survive in the exponent\cite{53} when one calculates the average $\left\langle A_{0}(\tau_{k})\ldots A_{0}(\tau_{1})\right\rangle_{S}$. Thus, in accordance with above considerations, applying the Baker-Hausdorf formula to each average of the type $\left\langle A_{0}(\tau_{k})\ldots A_{0}(\tau_{1})\right\rangle_{S}$ one can obtain following structure (notice, that $ k $ is even natural number everywhere )
\begin{align}\label{eq:series}
\begin{split}
\langle A_{0}(\tau_{k})A_{0}(\tau_{k-1})\ldots A_{0}(\tau_{2}) A_{0}(\tau_{1})\rangle_{S}= \\
e^{\frac{1}{2}\left\langle \left[\pm\varphi_{-}(\tau_{k})\mp \varphi_{-}(\tau_{k-1})\pm\ldots \pm \varphi_{-}(\tau_{2})\mp\varphi_{-}(\tau_{1}) \right]^2\right\rangle}\\
\times  cos \left[ f(\tau_{k}) - f(\tau_{k-1}) \right] \ldots cos \left[ f(\tau_{2}) - f(\tau_{1}) \right] 
 \end{split}
\end{align}
or, performing explicitly the squaring of the expression $[\pm\varphi_{-}(\tau_{k})\mp \varphi_{-}(\tau_{k-1})\pm\ldots \pm \varphi_{-}(\tau_{2})\mp\varphi_{-}(\tau_{1})]$ in the average $ \left\langle \left[\pm\varphi_{-}(\tau_{k})\mp \varphi_{-}(\tau_{k-1})\pm\ldots \pm \varphi_{-}(\tau_{2})\mp\varphi_{-}(\tau_{1}) \right]^2 \right\rangle $ in the exponent of Eq.(B10) one can obtain following structure
\begin{align}\label{eq:series}
\begin{split}
\langle A_{0}(\tau_{k})A_{0}(\tau_{k-1})\ldots A_{0}(\tau_{2}) A_{0}(\tau_{1})\rangle_{S}= \\
u(\tau_{k},\tau_{k-1},\ldots,\tau_{2},\tau_{1})C(\vert\tau_{k}-\tau_{k-1}\vert,\ldots,\vert\tau_{2}-\tau_{1}\vert). 
 \end{split}
\end{align}
Obviously, the expansion (B11) corresponds to the procedure where propagator $ \langle A_{0}(\tau_{k})\ldots A_{0}(\tau_{1})\rangle_{S} $  can be represented as the product of its "vertex" part ($ u(\tau_{k},\tau_{k-1},\ldots,\tau_{2},\tau_{1}) $ term) of $ k $-th order which describes all "crossing" diagrams (or propagators) while the rest is a product of $ k $ "free" propagators ($ C(\vert\tau_{k}-\tau_{k-1}\vert,\ldots,\vert\tau_{2}-\tau_{1}\vert)$ term) corresponding to a "linked cluster" expansion of $ k $ "non-crossing" diagrams. Further, one can check that mentioned "vertex" part $ u(\tau_{k},\tau_{k-1},\ldots,\tau_{2},\tau_{1}) $ from Eq.(B11) can always be arranged with respect to $ k $ time arguments in the following way
\begin{align}\label{eq:series}
\begin{split}
u(\tau_{k},\tau_{k-1},\ldots,\tau_{2},\tau_{1})=\prod_{l'=1}^{k-2} v(\tau_{l'},\tau_{l'+1};k)\\
=\prod_{l'=1}^{k-2}\left\lbrace \prod_{\begin{matrix}
m',j=1 \\
(m'\neq l'),(j\neq l'+1)
\end{matrix}}^{k}\frac{u(\tau_{l'}-\tau_{m'})}{u(\tau_{l'+1}-\tau_{j})}\right\rbrace . 
 \end{split}
\end{align}
Here we define following function $ v(\tau_{l'},\tau_{i};k) $ 
\begin{align}\label{eq:series}
\begin{split}
v(\tau_{l'},\tau_{l'+1};k)=\left\lbrace \prod_{\begin{matrix}
m',j=1 \\
(m'\neq l'),(j\neq l'+1)
\end{matrix}}^{k}\frac{u(\tau_{l'}-\tau_{m'})}{u(\tau_{l'+1}-\tau_{j})}\right\rbrace 
 \end{split}
\end{align}
with an evident property
\begin{align}\label{eq:series}
\begin{split}
v(\tau_{l'},\tau_{l'+1};k)=\frac{1}{v(\tau_{l'+1},\tau_{l'};k)}. 
 \end{split}
\end{align}
 In turn, the "linked cluster" (or "non-crossing diagrams") factor $C(\vert\tau_{k}-\tau_{k-1}\vert,\ldots,\vert\tau_{2}-\tau_{1}\vert)$ reads
\begin{align}\label{eq:series}
\begin{split}
C(\vert\tau_{k}-\tau_{k-1}\vert,\ldots,\vert\tau_{2}-\tau_{1}\vert)=u(\tau_{k} - \tau_{k-1}) \ldots u(\tau_{2} - \tau_{1})\\
\times cos \left[ f(\tau_{k}) - f(\tau_{k-1}) \right] \ldots cos \left[ f(\tau_{2}) - f(\tau_{1}) \right].
 \end{split}
\end{align}
In Eqs.(B12,B13) we have
\begin{align}\label{eq:series}
\begin{split}
u(\tau)=u(-\tau)=e^{-I(\tau)/g}
 \end{split}
\end{align}    
where $ I(\tau)=I(-\tau) $ is the symmetric "free" pair-correlator from Eq.(B8). 
Substitution of Eq.(B12) into Eq.(B11) gives us
\begin{align}\label{eq:series}
\begin{split}
\langle A_{0}(\tau_{k})A_{0}(\tau_{k-1})\ldots A_{0}(\tau_{2}) A_{0}(\tau_{1})\rangle_{S}= \\
= \left\lbrace \prod_{l'=1}^{k-2}v(\tau_{l'},\tau_{l'+1};k)\right\rbrace \times C(\vert\tau_{k}-\tau_{k-1}\vert,\ldots,\vert\tau_{2}-\tau_{1}\vert)
 \end{split}
\end{align}
The expression (B17) is, in fact, the result of Wick theorem application in order to extract absolute value from the average of the product  $ \langle A_{0}(\tau_{n})\ldots A_{0}(\tau_{1})\rangle_{S} $ of any $n $ operators $ A_{0}(\tau_{j}) $ being exponential in bosonic fields $ \varphi_{-}(\tau_{j}) $.
Now, using Eqs.(B11,B12,B17), one can write following identity
\begin{align}\label{eq:series}
\begin{split}
\int^{t}_{0} d \tau_{1}\int^{\tau_{1}}_{0} d \tau_{2}\ldots \int^{\tau_{n-2}}_{0} d \tau_{n-1}\int^{\tau_{n-1}}_{0} d \tau_{n} \\
\times \langle A_{0}(\tau_{1})A_{0}(\tau_{2})\ldots A_{0}(\tau_{n-1})A_{0}(\tau_{n})\rangle_{S} \\
=\int^{t}_{0} d \tau_{1}\int^{\tau_{1}}_{0} d \tau_{2}\ldots \int^{\tau_{n-2}}_{0} d \tau_{n-1}\int^{\tau_{n-1}}_{0} d \tau_{n} \\
\times u(\tau_{1},\tau_{2},\ldots,\tau_{n-1},\tau_{n})C(\vert\tau_{1}-\tau_{2}\vert,\ldots,\vert\tau_{n-1}-\tau_{n}\vert).
 \end{split}
\end{align} 
Now, on one hand, if in the l.h.s. of Eq.(B18) one exchanges two indices simultaneously in all pairs of "neighbouring" time arguments $ \tau_{r-1},\tau_{r} $ (notice on only one possible choice of such pairs in the product of given structure) as well as in the upper limits of corresponding time integrations, nothing will change since both sides of Eq.(B18) do not depend on any time arguments except $ t $ and since the average $ \langle A_{0}(\tau_{1})\ldots A_{0}(\tau_{n})\rangle_{S} $ is invariant under such exchange procedure by its definition (see above). This allows us to write
\begin{align}\label{eq:series}
\begin{split}
\int^{t}_{0} d \tau_{1}\int^{\tau_{1}}_{0} d \tau_{2}\ldots \int^{\tau_{n-2}}_{0} d \tau_{n-1}\int^{\tau_{n-1}}_{0} d \tau_{n} \\
\times \langle A_{0}(\tau_{1})A_{0}(\tau_{2})\ldots A_{0}(\tau_{n-1})A_{0}(\tau_{n})\rangle_{S} \\
=\int^{t}_{0} d \tau_{2}\int^{\tau_{2}}_{0} d \tau_{1}\ldots \int^{\tau_{n-3}}_{0} d \tau_{n}\int^{\tau_{n}}_{0} d \tau_{n-1} \\
\times \langle A_{0}(\tau_{2})A_{0}(\tau_{1})\ldots A_{0}(\tau_{n})A_{0}(\tau_{n-1})\rangle_{S} \\
=\int^{t}_{0} d \tau_{2}\int^{\tau_{2}}_{0} d \tau_{1}\ldots \int^{\tau_{n-3}}_{0} d \tau_{n}\int^{\tau_{n}}_{0} d \tau_{n-1} \\
\times \langle A_{0}(\tau_{1})A_{0}(\tau_{2})\ldots A_{0}(\tau_{n-1})A_{0}(\tau_{n})\rangle_{S}  .
 \end{split}
\end{align}
or 
\begin{align}\label{eq:series}
\begin{split}
\int^{t}_{0} d \tau_{2}\int^{\tau_{2}}_{0} d \tau_{1}\ldots \int^{\tau_{n-3}}_{0} d \tau_{n}\int^{\tau_{n}}_{0} d \tau_{n-1} \times \\
\langle A_{0}(\tau_{2})A_{0}(\tau_{1})\ldots A_{0}(\tau_{n})A_{0}(\tau_{n-1})\rangle_{S} \\
=\int^{t}_{0} d \tau_{2}\int^{\tau_{2}}_{0} d \tau_{1}\ldots \int^{\tau_{n-3}}_{0} d \tau_{n}\int^{\tau_{n}}_{0} d \tau_{n-1} \times\\
 u(\tau_{1},\tau_{2},\ldots,\tau_{n-1},\tau_{n})C(\vert\tau_{1}-\tau_{2}\vert,\ldots,\vert\tau_{n-1}-\tau_{n}\vert) .
 \end{split}
\end{align}  
 On the other hand, one can rewrite the l.h.s. of Eq.(B20) using Eq.(B17) in the form
\begin{align}\label{eq:series}
\begin{split}
\int^{t}_{0} d \tau_{2}\int^{\tau_{2}}_{0} d \tau_{1}\ldots \int^{\tau_{n-3}}_{0} d \tau_{n}\int^{\tau_{n}}_{0} d \tau_{n-1}\\ 
\times\langle A_{0}(\tau_{2})A_{0}(\tau_{1})\ldots A_{0}(\tau_{n})A_{0}(\tau_{n-1})\rangle_{S}= \\
\int^{t}_{0} d \tau_{2}\int^{\tau_{2}}_{0} d \tau_{1}v(\tau_{2},\tau_{1};n)\ldots \\ 
\ldots \int^{\tau_{n-3}}_{0} d \tau_{n}\int^{\tau_{n}}_{0} d \tau_{n-1}v(\tau_{n},\tau_{n-1};n) \\
\times C(\vert\tau_{1}-\tau_{2}\vert,\ldots,\vert\tau_{n-1}-\tau_{n}\vert) .
 \end{split}
\end{align}  
which in turn can be rewritten using property (B14) as
\begin{align}\label{eq:series}
\begin{split}
\int^{t}_{0} d \tau_{2}\int^{\tau_{2}}_{0} d \tau_{1}v(\tau_{2},\tau_{1};n)\ldots \\ 
\ldots \int^{\tau_{n-3}}_{0} d \tau_{n}\int^{\tau_{n}}_{0} d \tau_{n-1}v(\tau_{n},\tau_{n-1};n) \\
\times C(\vert\tau_{1}-\tau_{2}\vert,\ldots,\vert\tau_{n-1}-\tau_{n}\vert) \\
=\int^{t}_{0} d \tau_{2}\int^{\tau_{2}}_{0} d \tau_{1}\frac{1}{v(\tau_{1},\tau_{2};n)}\ldots \\ 
\ldots \int^{\tau_{n-3}}_{0} d \tau_{n}\int^{\tau_{n}}_{0} d \tau_{n-1}\frac{1}{v(\tau_{n-1},\tau_{n};n)}\\
\times C(\vert\tau_{1}-\tau_{2}\vert,\ldots,\vert\tau_{n-1}-\tau_{n}\vert) .
 \end{split}
\end{align} 
Now, substituting Eq.(B22) into the r.h.s. of Eq.(B20) and using definition Eq.(B12) one can perform Eq.(B20) in the form
\begin{align}\label{eq:series}
\begin{split}
\int^{t}_{0} d \tau_{2}\int^{\tau_{2}}_{0} d \tau_{1}\ldots \int^{\tau_{n-3}}_{0} d \tau_{n}\int^{\tau_{n}}_{0} d \tau_{n-1} \\
\times \langle A_{0}(\tau_{2})A_{0}(\tau_{1})\ldots A_{0}(\tau_{n})A_{0}(\tau_{n-1})\rangle_{S} \\
=\int^{t}_{0} d \tau_{2}\int^{\tau_{2}}_{0} d \tau_{1}\ldots \int^{\tau_{n-3}}_{0} d \tau_{n}\int^{\tau_{n}}_{0} d \tau_{n-1} \\
\times \frac{1}{u(\tau_{1},\tau_{2},\ldots,\tau_{n-1},\tau_{n})}C(\vert\tau_{1}-\tau_{2}\vert,\ldots,\vert\tau_{n-1}-\tau_{n}\vert).
 \end{split}
\end{align}   
 Finally, changing back indices in all pairs of "neighbouring" time arguments $ \tau_{r-1},\tau_{r}\rightarrow \tau_{r},\tau_{r-1}$ in Eq.(23) with respect to symmetry property of Eq.(B19)and comparing the left- and right-hand sides of Eqs.(B18,B20,B23) one can conclude that
\begin{align}\label{eq:series}
\begin{split}
\int^{t}_{0} d \tau_{1}\int^{\tau_{1}}_{0} d \tau_{2}\ldots \int^{\tau_{n-2}}_{0} d \tau_{n-1}\int^{\tau_{n-1}}_{0} d \tau_{n} \\
\times \left\lbrace \textbf{K}_{\lbrace n\rbrace}-{\textbf{K}_{\lbrace n\rbrace}}^{-1}\right\rbrace  C(\vert\tau_{1}-\tau_{2}\vert,\ldots,\vert\tau_{n-1}-\tau_{n}\vert) =0.
 \end{split}
\end{align} 
In Eq.(B24) the "kernel" function
\begin{align}\label{eq:series}
\begin{split}
\textbf{K}_{\lbrace n\rbrace}=u(\tau_{1},\tau_{2},\ldots,\tau_{n-1},\tau_{n})
 \end{split}
\end{align}
is a kind of generalized function which can act on any function $ f(\tau_{1},\ldots,\tau_{n}) $ only under $ n $-fold time-integration over $\tau_{1},.., \tau_{n} $ with the following evident property
\begin{align}\label{eq:series}
\begin{split}
\textbf{K}_{\lbrace n\rbrace}{\textbf{K}_{\lbrace n\rbrace}}^{-1}={\textbf{K}_{\lbrace n\rbrace}}^{-1}\textbf{K}_{\lbrace n\rbrace}=1.
 \end{split}
\end{align} 
Obviously, from Eq.(B24) it follows that
\begin{align}\label{eq:series}
\begin{split}
\int^{t}_{0} d \tau_{1}\int^{\tau_{1}}_{0} d \tau_{2}\ldots \int^{\tau_{n-2}}_{0} d \tau_{n-1}\int^{\tau_{n-1}}_{0} d \tau_{n} \\
\textbf{K}_{\lbrace n\rbrace} \times C(\vert\tau_{1}-\tau_{2}\vert,\ldots,\vert\tau_{n-1}-\tau_{n}\vert) \\
= \int^{t}_{0} d \tau_{1}\int^{\tau_{1}}_{0} d \tau_{2}\ldots \int^{\tau_{n-2}}_{0} d \tau_{n-1}\int^{\tau_{n-1}}_{0} d \tau_{n} \\
{\textbf{K}_{\lbrace n\rbrace}}^{-1}\times C(\vert\tau_{1}-\tau_{2}\vert,\ldots,\vert\tau_{n-1}-\tau_{n}\vert).
 \end{split}
\end{align} 
Then since by definition $ C(\vert\tau_{1}-\tau_{2}\vert,\ldots,\vert\tau_{n-1}-\tau_{n}\vert)> 0 $ (see Eqs.(B15,B16)), it follows that the only possibility for kernel generalized function of Eq.(B25) to fulfil both Eq.(B26) and Eq.(B27) simultaneously is
\begin{align}\label{eq:series}
\begin{split}
\textbf{K}_{\lbrace n\rbrace}={\textbf{K}_{\lbrace n\rbrace}}^{-1}=\textbf{1}_{\lbrace n\rbrace}=\tilde{1}(\tau_{1},\tau_{2},\ldots,\tau_{n-1},\tau_{n}).
 \end{split}
\end{align} 
Here by means of Eq.(B28) I defined the $ n $ -dimensional "generalized unit function": $ \tilde{1}(\tau_{1},\ldots,\tau_{n}) $ which is a sort of generalized function (or operator) being "unit" in the sense that the result of expression $ \textbf{1}_{\lbrace n\rbrace}C(\vert\tau_{1}-\tau_{2}\vert,\ldots,\vert\tau_{n-1}-\tau_{n}\vert) $ after all the integrations over all $ \tau_{1},\ldots,\tau_{n} $ time arguments in Eq.(B27) will be the same as if one would integrate only the function $ C(\vert\tau_{1}-\tau_{2}\vert,\ldots,\vert\tau_{n-1}-\tau_{n}\vert)$  over those time arguments   
\begin{align}\label{eq:series}
\begin{split}
\int^{t}_{0} d \tau_{1}\int^{\tau_{1}}_{0} d \tau_{2}\ldots \int^{\tau_{n-2}}_{0} d \tau_{n-1}\int^{\tau_{n-1}}_{0} d \tau_{n} \\
\times\textbf{K}_{\lbrace n\rbrace}C(\vert\tau_{1}-\tau_{2}\vert,\ldots,\vert\tau_{n-1}-\tau_{n}\vert) \\
=\int^{t}_{0} d \tau_{1}\int^{\tau_{1}}_{0} d \tau_{2}\ldots \int^{\tau_{n-2}}_{0} d \tau_{n-1}\int^{\tau_{n-1}}_{0} d \tau_{n} \\
\times C(\vert\tau_{1}-\tau_{2}\vert,\ldots,\vert\tau_{n-1}-\tau_{n}\vert).
 \end{split}
\end{align}
In turn, the latter equality Eq.(B29) means that, without the loss of generality, one can perform Eq.(B18) simply as
\begin{align}\label{eq:series}
\begin{split}
\int^{t}_{0} d \tau_{1}\int^{\tau_{1}}_{0} d \tau_{2}\ldots \int^{\tau_{n-2}}_{0} d \tau_{n-1}\int^{\tau_{n-1}}_{0} d \tau_{n} \\
\times \langle A_{0}(\tau_{1})A_{0}(\tau_{2})\ldots A_{0}(\tau_{n-1})A_{0}(\tau_{n})\rangle_{S} \\
=\int^{t}_{0} d \tau_{1}\int^{\tau_{1}}_{0} d \tau_{2}\ldots \int^{\tau_{n-2}}_{0} d \tau_{n-1}\int^{\tau_{n-1}}_{0} d \tau_{n} \\
\times C(\vert\tau_{1}-\tau_{2}\vert,\ldots,\vert\tau_{n-1}-\tau_{n}\vert).
 \end{split}
\end{align}
Finally, evident properties (see Eqs.(B15,B16))
\begin{align}\label{eq:series}
\begin{split}
 \langle A_{0}(\tau_{1})A_{0}(\tau_{2})\rangle_{S}= C(\vert\tau_{1}-\tau_{2}\vert)
 \end{split}
\end{align}
and
\begin{align}\label{eq:series}
\begin{split}
C(\vert\tau_{1}-\tau_{2}\vert,\ldots,\vert\tau_{n-1}-\tau_{n}\vert) \\
=C(\vert\tau_{1}-\tau_{2}\vert)\ldots C(\vert\tau_{n-1}-\tau_{n}\vert)
 \end{split}
\end{align}
allow for the \textit{exact} factorization of the average under the integrals in the l.h.s. of Eq.(B30) on a product of $ n/2 $ pair-correlators( recall that $ n $ is an arbitrary even number everywhere)
\begin{align}\label{eq:series}
\begin{split}
\int^{t}_{0} d \tau_{1}\int^{\tau_{1}}_{0} d \tau_{2}\ldots \int^{\tau_{n-2}}_{0} d \tau_{n-1}\int^{\tau_{n-1}}_{0} d \tau_{n} \\
\times \langle A_{0}(\tau_{1})A_{0}(\tau_{2})\ldots A_{0}(\tau_{n-1})A_{0}(\tau_{n})\rangle_{S} \\
=\int^{t}_{0} d \tau_{1}\int^{\tau_{1}}_{0} d \tau_{2}\ldots \int^{\tau_{n-2}}_{0} d \tau_{n-1}\int^{\tau_{n-1}}_{0} d \tau_{n} \\
\times\langle A_{0}(\tau_{1}) A_{0}(\tau_{2})\rangle_{S}\ldots \langle A_{0}(\tau_{n-1}) A_{0}(\tau_{n})\rangle_{S} .
 \end{split}
\end{align}

Now, it is time to consider all "non-time ordered" contributions to the Keldysh contour-ordered time-integrals from Eq.(B1). In particular, taking closer look on averages of the kind $\langle \mathcal{T}_K A_{0}(\tau_{k})\ldots A_{0}(\tau_{1})A_{0}(\tau'_{n-k})\ldots A_{0}(\tau'_{1})\rangle_{S}$ under time- integrals over $ \tau_{j} $ ($ j'=1,..,k $) and $ \tau'_{j'} $ ($ j'=1,..,(n-k) $) in terms from the T-exponent expansion in r.h.s. of Eq.(8) from the main text, one can notice that application of Keldysh-contour ordering procedure to this average with respect to Eq.(B33) derived above - leads to the appearance of all possible "mixed" pair-correlators of the kind $ \int_{0}^{t} \int_{0}^{\tau_{j+1}} d \tau_{j}d \tau'_{j'} \langle A_{0}(\tau_{j})A_{0}(\tau'_{j'})\rangle_{S} $ in corresponding factorization formulas of Eq(B33). The latter involve operators $ A_{0}(\tau_{j}) $ and $ A_{0}(\tau'_{j'}) $ from different branches of Keldysh contour (or, alternatively, from both time- and anti-time-ordered sequences of such operators in the r.h.s. of Eq.(8)). At first glance, the correlators of such type should "break" the sequence of time integrations in the l.h.s. of Eqs.(B1,B33) because, for example, for the "non-time-oredered" average of the kind $\int_{0}^{t} \int_{0}^{\tau_{j+1}} d \tau_{j}d \tau'_{j'}\langle A_{0}(\tau_{j})A_{0}(\tau'_{j'})\rangle_{S} $ two corresponded integrations (over $ \tau_{j} $ and over $ \tau'_{j'} $) to appear in r.h.s. of Eq.(B1) are disconnected. However, this obstacle can be circumvented by decomposing each contribution of the kind $ \int_{0}^{t} \int_{0}^{\tau_{j+1}} d \tau_{j}d \tau'_{j'} $ (in the expansion of r.h.s. of Eq.(8)) on its "time-" and "anti-time" -ordered parts (with respect to the cases $ \tau_{j}>\tau'_{j'}$ and $ \tau_{j}<\tau'_{j'}$, correspondingly) and, then, by "assigning" each (anti-)time-ordered "part" of this double integral to the (anti-)time-ordered sequence of integrations in Eqs.(B1,B33) in order to "restore" the "broken" sequence  of time-ordered integrations to the $ n $-fold integral over $ n $ time variables $ \tau_{1},\ldots \tau_{n} $. 
Obviously, as the result, all sequences of $ n $ time-ordered integrations being obtained in such a way will be the same as one in the r.h.s. of Eq.(B33). Hence, afterwards, one will need only to count all these sequences properly, extracting a correct combinatoric pre-factor in front of the sequence of $ n $ time-integrations similar to one from the r.h.s. of Eq.(B33). 
Applying this procedure, one can easily convince that
\begin{align}\label{eq:comb}
\begin{split}
\int^{t}_{0} d \tau_{1}\int^{\tau_{1}}_{0} d \tau_{2}\ldots \int^{\tau_{k-2}}_{0} d \tau_{k-1}\int^{\tau_{k-1}}_{0} d \tau_{k}\times  \\
 \int^{t}_{0} d \tau'_{1}\int^{\tau'_{1}}_{0} d \tau'_{2}\ldots \int^{\tau'_{(n-k)-2}}_{0} d \tau'_{n-k-1}\int^{\tau'_{(n-k)-1}}_{0} d \tau'_{n-k}\\
\times \langle \mathcal{T}_K A_{0}(\tau_{k})\ldots A_{0}(\tau_{1}) A_{0}(\tau'_{n-k})\ldots A_{0}(\tau'_{1})\rangle_{S} \\
=\textbf{D}_{(n)}\times \int^{t}_{0} d \tau_{1}\int^{\tau_{1}}_{0} d \tau_{2}\langle A_{0}(\tau_{1})A_{0}(\tau_{2})\rangle_{S}\ldots \\
\ldots \int^{\tau_{n-2}}_{0} d \tau_{n-1} \int^{\tau_{n-1}}_{0} d \tau_{n}\langle A_{0}(\tau_{n-1})A_{0}(\tau_{n})\rangle_{S}. 
\end{split}
\end{align}
where
\begin{align}\label{eq:rel}
\begin{split}
\textbf{D}_{(n)}=\sum_{j=0}^{k}C_{j}^{k}\sum_{j'=0}^{n-k}C_{j'}^{n-k}=2^{k}\cdot 2^{n-k}=2^{n}
\end{split}
\end{align} 
is the desired combinatoric pre-factor. This factor gives us the number of ways in which one could "compose" all "mixed" correlators from two sequences of time arguments $ \tau_{1}\ldots \tau_{k}$ and $ \tau'_{1}\ldots \tau'_{n-k}$. In Eq.(B35) I  used a usual binomial formula $\sum_{j=0}^{k}C_{j}^{k}=2^{k} $ for binomial coefficients $ C_{j}^{k}=\frac{k!}{j!(k-j)!} $ and $ C_{j'}^{k}=\frac{(n-k)!}{j'!(n-k-j')!} $ which counts the numbers of different ways one could put $ j $ and $ j' $ "plugs" into the sequences of $ k $ "time-ordered" and $ n-k $ "anti-time ordered" integrations, correspondingly, in the r.h.s. of Eq.(B33). 
 At last, taking into account that in according with Eqs.(B6,B11) $ \langle A_{0}(\tau_{l+1}) A_{0}(\tau_{l})\rangle_{S}=\langle A_{0}(\tau_{l}) A_{0}(\tau_{l+1})\rangle_{S} $  and also recalling the fact that everywhere in the above formulas $ n=2m $ (i.e. $ n $ is even natural number), with the help of the obvious property for $ m=n/2 $-fold double integral
\begin{align}\label{eq:rel}
\begin{split}
\int^{t}_{0} d \tau_{1}\int^{\tau_{1}}_{0} d \tau_{2}\langle A_{0}(\tau_{1})A_{0}(\tau_{2})\rangle_{S}\ldots \\
\ldots \int^{\tau_{n-2}}_{0} d \tau_{n-1} \int^{\tau_{n-1}}_{0} d \tau_{n}\langle A_{0}(\tau_{n-1})A_{0}(\tau_{n})\rangle_{S}\\ 
= \frac{1}{(n/2)!} \prod_{l=1}^{n/2} \left\lbrace  \int^{t}_{0} d \tau_{l+1}\int^{\tau_{l+1}}_{0} d \tau_{l}\langle A_{0}(\tau_{l+1}) A_{0}(\tau_{l})\rangle_{S}\right\rbrace \\
= \frac{1}{(n/2)!} \prod_{l=1}^{n/2} \left\lbrace \frac{1}{2} \int^{t}_{0} d \tau_{l+1}\int^{t}_{0} d \tau_{l}\langle A_{0}(\tau_{l+1}) A_{0}(\tau_{l})\rangle_{S}\right\rbrace 
\end{split}
\end{align}
and combining Eqs.(B34-B36) one can easily obtain following \textit{exact} equality
\begin{align}\label{eq:comb}
\begin{split}
\int^{t}_{0} d \tau_{1}\int^{\tau_{1}}_{0} d \tau_{2}\ldots \int^{\tau_{k-2}}_{0} d \tau_{k-1}\int^{\tau_{k-1}}_{0} d \tau_{k}\times  \\
 \int^{t}_{0} d \tau'_{1}\int^{\tau'_{1}}_{0} d \tau'_{2}\ldots \int^{\tau'_{(n-k)-2}}_{0} d \tau'_{n-k-1}\int^{\tau'_{(n-k)-1}}_{0} d \tau'_{n-k}\\
\times \langle \mathcal{T}_K A_{0}(\tau_{k})\ldots A_{0}(\tau_{1}) A_{0}(\tau'_{n-k})\ldots A_{0}(\tau'_{1})\rangle_{S} \\
=\frac{2^{n}}{(n/2)!} \prod_{l=1}^{n/2} \left\lbrace \frac{1}{2} \int^{t}_{0} d \tau_{l+1}\int^{t}_{0} d \tau_{l}\langle A_{0}(\tau_{l+1}) A_{0}(\tau_{l})\rangle_{S} \right\rbrace \\
= \frac{1}{(n/2)!} \prod_{l=1}^{n/2} \left\lbrace 2 \int^{t}_{0} d \tau_{l+1}\int^{t}_{0} d \tau_{l}\langle A_{0}(\tau_{l+1}) A_{0}(\tau_{l})\rangle_{S} \right\rbrace . 
\end{split}
\end{align}
which states S-Lemma (since the r.h.s. of Eq.(B37) coincides with corresponding Eq.(16) from the main text). \textit{Therefore, S-Lemma is proven.} $ \blacksquare $
The proof of the equality (B37) is, in fact, a basic claim of this section: I proved here that \textit{all} the "crossing" diagrams in the expansion (B9) \textit{do not affect} the result of corresponding time integrations.  In other words, here it has been obtained that a so-called "linked cluster approximation" for Luttinger liquid tunnel junction represents an \textit{exact} procedure, which takes place when one integrates averages $ \langle A_{0}(\tau_{1})A_{0}(\tau_{2})\ldots A_{0}(\tau_{n-1})A_{0}(\tau_{n})\rangle_{S} $ over all available time arguments and, obviously, this claim remains valid in \textit{all} orders of perturbation theory in the tunnel coupling constant.

Now, substituting Eq.(B37) into the r.h.s. of Eq.(B1) one can \textit{exactly} transform expression for $ \tilde{Z}_{12(21)}(t) $ to the form 
\begin{align}\label{eq:comb}
\begin{split}
\tilde{Z}_{12(21)}(t)=1+\sum_{n=2}^{even}  
\sum_{j,k=0}^{n} C^{n}_{k} 
 e^{i \vartheta_{g} k}e^{-i \vartheta_{g}(n/2-k)} \\
    \times C^{n}_{j}
   (i\tilde{\lambda}_{1(2)})^{(j+k)} 
    (-i\tilde{\lambda}_{2(1)})^{(n-(j+k))}  \\
     \times \frac{1}{(n/2)!}  \prod_{l=1}^{n/2} \left\lbrace 2\int^{t}_{0} d \tau_{l+1}\int^{t}_{0} d \tau_{l}
      \langle A_{0}(\tau_{l+1}) A_{0}(\tau_{l})\rangle_{S}\right\rbrace . 
\end{split}
\end{align}
Obviously, from all the above it follows that the expansion in the r.h.s. of Eq.(B38) automatically takes care about all possible combinations constructed  from the correlators \textit{of all orders} which can appear in the basic power expansion of Eq.(8) from the main text. Thus, the result of Eq.(B38) remains valid in \textit{all} orders in $ n $ giving rise to the non-perturbative calculation of $ \tilde{Z}_{12(21)}(t) $. Indeed, using the fact that in Eq.(B38) $ n=2m  $, ($ m=1,2,3.. $) and applying twice a binomial formula $ (x+y)^{m}=\sum_{k=1}^{m} C^{m}_{k}x^{k}y^{m-k} $ one obtains from Eq.(B38) following \textit{exact} decomposition
\begin{align}\label{eq:series}
\begin{split}
\tilde{Z}_{12(21)}(t)=1+\sum_{m=1}^{\infty}  
(-1)^{m} \left( \tilde{\lambda}_{1(2)}-\tilde{\lambda}_{2(1)}\right)^{m} \\
\times \left\lbrace  \tilde{\lambda}_{1(2)}e^{i \vartheta_{g}}-\tilde{\lambda}_{2(1)}e^{-i \vartheta_{g}} \right]^{m} \\
\times \frac{1}{m!} \left[2\int^{t}_{0} d \tau_{1}\int^{t}_{0} d \tau_{2}
      \langle A_{0}(\tau_{1}) A_{0}(\tau_{2})\rangle_{S} \right\rbrace^{m} ,
 \end{split}
\end{align}
which, in turn, can be re-exponentiated exactly to the obvious compact form
\begin{eqnarray}\label{eq:reex}
\tilde{Z}_{12(21)}(t)=\exp \left\lbrace - \textit{F}_{12(21)}(t) \right\rbrace 
\end{eqnarray}
where for the function $ \textit{F}_{12}(t)=\textit{F}^{\ast}_{21}(t) $ one has
\begin{eqnarray}\label{eq:F}
\nonumber
\textit{F}_{12(21)}(t)=\left( \tilde{\lambda}_{1(2)}-\tilde{\lambda}_{2(1)}\right)\left[ \tilde{\lambda}_{1(2)}e^{i \vartheta_{g}}-\tilde{\lambda}_{2(1)}e^{-i \vartheta_{g}} \right]\nonumber\\
\times \left\lbrace 2\int^{t}_{0} d \tau_{1}\int^{t}_{0} d \tau_{2}\langle A_{0}(\tau_{1}) A_{0}(\tau_{2})\rangle_{S} \right\rbrace .
\nonumber
\end{eqnarray}
\begin{equation}
\end{equation}
Remarkably, using the symmetry property of Eq.(B39) for two-fold time-integral one can also rewrite the resulting formula of Eq.(B41) in its more compact (and Keldysh contour-ordered) form
\begin{eqnarray}\label{eq:F}
\nonumber
\textit{F}_{12(21)}(t)=\left( \tilde{\lambda}_{1(2)}-\tilde{\lambda}_{2(1)}\right)\left[ \tilde{\lambda}_{1(2)}e^{i \vartheta_{g}}-\tilde{\lambda}_{2(1)}e^{-i \vartheta_{g}} \right]\nonumber\\
\times \left\lbrace \int \int_{\textit{C}_{K}} d \tau_{1} d \tau_{2}\langle A_{0}(\tau_{1}) A_{0}(\tau_{2})\rangle_{S} \right\rbrace .
\nonumber
\end{eqnarray}
\begin{equation}
\end{equation}  
where the integrations over $ \tau_{1}$, $ \tau_{2}$ in the r.h.s. of Eq.(B42) are taken along the complex Keldysh contour $ \mathcal{C}_K\in (0-i\beta;t) $ (one can compare it with the Eqs.(13-15) from the main text).
Evidently, \textit{exact} equations (B40-B42) have been derived here - do coincide with the equations (13-15) from the main text and, hence, state the claim of the \textit{S-Theorem}. \textit{Therefore, S-Theorem is proven.} $ \blacksquare $


\begin{references}

%gen_qmeas_contr

\bibitem{1} H.Wiseman, G.J.Milburn, {\em Quantum Measurement and
Control}, Cambridge University Press, Cambridge (2010).
%qnoise_meas_ampl_rev
\bibitem{2} A.A.Clerk, M.H.Devoret, S.M.Girvin, F.Marquardt,
R.J.Schoelkopf, {\em Rev.Mod.Phys.} {\bf 82}, 1155 (2010).
%exp_singl_shot_readout_spin_silic
\bibitem{3} A.Morello, J.J.Pla, F.A.Zwanenburg, K.W.Chan, K.Y.Tan, H.Huebl, M.Möttönen, C.D.Nugroho, C.Yang, J.A.van
Donkelaar et al., {\em Nature}, London {\bf 467}, 687 (2010).
%quant_control_dec_mol_struct
\bibitem{4} M.Shapiro and P.Brumer, {\em Quantum Control of Molecular
Processes}, Wiley, Weinheim, (2012).

%cont_qmeas_dqd

\bibitem{5} A.N.Korotkov, {\em Phys.Rev.B}, {\bf 60}, 5737 (1999).
%qbit_cont_meas_theor
\bibitem{6} A.N.Korotkov, {\em Phys.Rev.B}, {\bf 63}, 115403 (2001).

%interaction_problems

\bibitem{7} M.Sciro, A.Mitra, {\em Phys.Rev.Lett.}, {\bf 112}, 246401 (2014).
%nonequil_Lutt
\bibitem{8} D.B.Gutman, Y.Gefen, A.D.Mirlin {\em Phys.Rev.B}, {\bf 81}, 085436 (2010).

%main_qdet_theor

\bibitem{9} S.A.Gurvitz, {\em Phys. Rev.B}, {\bf 56}, 15215 (1997).
%main_q_det_theor
\bibitem{10} D.V.Averin and E.V.Sukhorukov, {\em Phys.Rev.Lett.}, {\bf 95}, 126803
(2005).

%dqd_exp_platforms

%main_exp_qpc_qdet
\bibitem{11} J.Gorman, D.G.Hasko, D.A.Williams, {\em Phys.Rev.Lett.}, {\bf 95}, 090502, (2005).
%main_exp_qpc_qdet_microwave 
\bibitem{12} K.D.Petersson, J.R.Petta, H.Lu, A.C.Gossard, {\em Phys.Rev.Lett.}, {\bf 105}, 246804 (2010).
%exp_manip_charg_dqd_readout
\bibitem{13} J.M.Elzerman, R.Hanson, L.H.Willems van Beveren, B.Witkamp, L.M.K.Vandersypen, L.P.Kouwenhoven,
{\em Nature}, London {\em 430}, 431 (2004); N.P.Oxtoby, H.M.Wiseman, H.-B.Sun, {\em Phys.Rev.B}, {\bf 74},
045328 (2006). 
%SET_qdet_charg_qbit
\bibitem{14}J.R.Petta, A.C.Johnson, J.M.Taylor, E.A.Laird, A.Yacoby, M.D.Lukin, C.M.Marcus, M.P.Hanson, A.C.Gossard, {\em Science}, {\bf 309}, 2180 (2005); H.-O. Li, G.C.M.Xiao, J.You, D.Wei, T.Tu, G.-C. Guo, H.-W. Jiang, G.-P. Guo, {\em J.Appl.Phys.}, {\bf 116}, 174504 (2014).

%noneq_Lutt_theor
\bibitem{15} S. Ngo Dinh, D.A. Bagrets, A.D. Mirlin, {\em Phys.Rev.B}, {\bf 81}, 081306 (2010).
%nonequil_Lutt_eff_charge_frac
\bibitem{16} I.Safi, H.J.Schulz, {\em Phys.Rev.B}, {\bf 52}, R17040(R), (1995).
%nonequil_Lutt_eff_coupl_to_electrodes
\bibitem{17} Ya.M.Blanter, F.W.J.Hekking, M.Buettiker, {\em Phys.Rev.Lett.}, {\bf 81}, 1925, (1998).

%exp_qcontrol_meas

%first_meas_coul_bl
\bibitem{18} M.Field, C.G.Smith, M.Pepper, D.A.Ritchie, J.E.F.Frost, G.A.C.Jones, D.G.Hasko, {\em Phys.Rev.Lett.}, {\bf 70}, 1311 (1993).
%exp_state_cond_rabi_quadr_qd
\bibitem{19} D.R.Ward, D.Kim, D.E.Savage, M.G.Lagally, R.H.Foote, M.Friesen, S.N.Coppersmith, M.A.Eriksson, {\em npj Quant.Inf.}, {\bf 2}, 16032 (2016).
%cont_q_feedb_theor_qbit
\bibitem{20} Q.Zhang, R.Ruskov, A.N.Korotkov, {\em Phys.Rev.B}, {\bf 72}, 245322 (2005).
%exp_q_feedb_contr_rabi
\bibitem{21} R.Vijay, C.Macklin, D.H.Slichter, S.J.Weber, K.W.Murch, R.Naik, A.N.Korotkov, I.Siddiqi, {\em Nature} (London) {\bf 490}, 77 (2012).

%benefits_weak

%Lutt_weak_meas_theor
\bibitem{22} A.Bednorz, W.Belzig, {\em Phys.Rev.Lett.}, {\bf 105}, 106803 (2010); A.Bednorz, W.Belzig, {\em Phys.Rev.Lett.}, {\bf 81}, 125112, (2010).
%q_comp_q_mem_schem_spin-spin_weak_int
\bibitem{23} S.Kagami, Y.Shikano, K.Asahi, {\em J.Phys.E}, {\bf 43}, 761 (2011).
%part_meas_backact_noncl_weak_val_superc_circ
\bibitem{24} J.P.Groen, D.Ristè, L.Tornberg, J.Cramer, P.C.deGroot,
T.Picot, G.Johansson, L.DiCarlo, {\em Phys.Rev.Lett.}, {\bf 111},
090506 (2013); C.Meyer zu Rheda, G.Haack, A.Romito, {\em Phys.Rev.B}, {\bf 90},
155438 (2014).
%exp_phot_spin_hall_eff_weak_meas
\bibitem{25} O.Hosten, P.Kwiat, {\em Science}, {\bf 319}, 787 (2008).
%exp_sagn_interf_weak_val
\bibitem{26} P.B.Dixon, D.J.Starling, A.N.Jordan, J.C.Howell, {\em Phys.Rev.Lett.}, {\bf 102}, 173601 (2009).
%q_state_discr_meas_weak
\bibitem{27} A.N.Jordan, J.Tollaksen, J.E.Troupe, J.Dressel, Y.Aharonov, {\em Quant.Stud.:Math.Found.}, {\bf 2}, 5 (2015).
%q_state_discr_prot_weak
\bibitem{28} O.Zilberberg, A.Romito, D.J.Starling, G.A.Howland, C.J.Broadbent, J.C.Howell, Y. Gefen, {\em Phys.Rev.Lett.}, {\bf 110},170405 (2013).
%quant_meas_q_traj_weak_meas
\bibitem{29} S.J.Weber, A.Chantasri, J.Dressel, A.N.Jordan, K.W.Murch, I.Siddiqi, {\em Nature} (London), {\bf 511}, 570 (2014).

%benefits_gen

%nmr_techn_q_contr_comp
\bibitem{30} L.M.K.Vandersypen, I.L.Chuang, {\em Rev.Mod.Phys.}, {\bf 76},
1037 (2005).
%majoran_calc_dec
\bibitem{31} D.Aasen, M.Hell, R.V.Mishmash, A.Higginbotham, J.Danon, M.Leijnse, T.S.Jespersen, J.A.Folk, C.M.Marcus, K.Flensberg et al., {\em Phys.Rev.X}, {\bf 6}, 031016 (2016).
%exp_echo_meas_chrg_qbit
\bibitem{32} Z.Shi, C.B.Simmons, D.R.Ward, J.R.Prance, R.T.Mohr, T.S.Koh, J.K.Gamble, X.Wu, D.E.Savage, M.G.Lagally et al., {\em Phys.Rev.B}, {\bf 88}, 075416 (2013).

%exp_dqd_preparation platforms

%exp_ultraf_charg_contr_dqd
\bibitem{33} G.Cao, H.-O.Li, T.Tu, L.Wang, C.Zhou, M.Xiao, G.-C.Guo, H.-W.Jiang, G.-P.Guo, {\em Nat.Commun.}, {\bf 4}, 1401 (2013).
%exp_microw_op_charg_qbit
\bibitem{34} D.Kim, D.R.Ward, C.B.Simmons, J.K.Gamble, R.Blume-
Kohout, E.Nielsen, D.E.Savage, M.G.Lagally, M.Friesen, S.N.Coppersmith, M.A.Eriksson, {\em Nat. Nanotechnol.}, {\bf 10(3)}, 243 (2015).
%exp_few_el_dqd_microw_rad_induced_manip_two_cdet
\bibitem{35} J.M.Elzerman, R.Hanson, J.S.Greidanus, L.H.Willems van
Beveren, S.De Franceschi, L.M.K.Vandersypen, S.Tarucha, L.P.Kouwenhoven, {\em Phys.Rev.B}, {\bf 67}, 161308 (2003).
%exp_manip_singl_charg_dqd
\bibitem{36} J.R.Petta, A.C.Johnson, C.M.Marcus, M.P.Hanson, A.C.Gossard, {\em Phys.Rev.Lett.}, {\bf 93}, 186802 (2004).

%dec_control_platforms

%control_dec
\bibitem{37} A.J.Leggett, S.Chakravarty, A.T.Dorsey, M.P.A.Fisher, A.Garg, W.Zwerger, {\em Rev.Mod.Phys.}, {\bf 59}, 1 (1987).
\bibitem{38} U.Weiss, {\em Quantum Dissipative Systems}, World Scientiﬁc,Singapore, (2012).
%control_dec
\bibitem{39} M.P.A.Fisher, L.I.Glazman, {\em Mesoscopic Electron Transport}, edited by L. L. Sohn, L. P. Kouwenhoven, G. Schön, NATO Advanced Study Institute, Springer, The Netherlands, (1997).
%quant_lim_squid
\bibitem{40} A.A.Clerk, {\em Phys.Rev.Lett.}, {\bf 96}, 056801 (2006).

%theor_backgr

%main_wich_path
\bibitem{41} I.L.Aleiner, N.S.Wingreen, Y.Meir, {\em Phys.Rev.Lett.}, {\bf 79}, 3740 (1997).

%Lutt_exp_backgr

%Lutt_exp
\bibitem{42} T.Li, P.Wang, H.Fu, L.Du, K.A.Schreiber, X.Mu, X.Liu, G.Sullivan, G.A.Csáthy, X.Lin, R.-R.Du, {\em Phys.Rev.Lett.}, {\bf 115}, 136804 (2015).
\bibitem{43} E.Levy, I.Sternfeld, M.Eshkol, M.Karpovski, B.Dwir, A.Rudra, E.Kapon, Y.Oreg, A.Palevski, {\em Phys.Rev.B}, {\bf 85}, 045315 (2012); J.Dai, J.Li, H.Zeng, X.Cui, {\em Appl.Phys.Lett.}, {\bf 94}, 093114 (2009).
\bibitem{44} H.Ishii, H.Kataura, H.Shiozawa, H.Yoshioka, H.Otsubo, Y.Takayama, T.Miyahara, S.Suzuki, Y.Achiba, M.Nakatake,
T.Narimura, M.Higashiguchi, K.Shimada, H.Namatame, M.Taniguchi, {\em Nature}, (London) {\bf 426}, 540 (2003).

%Lutt_theor_backgr

%main_KF_eff
\bibitem{45} C.L.Kane, M.P.A.Fisher, {\em Phys.Rev.Lett.}, {\bf 68}, 1220 (1992).
%main_Lutt_bound
\bibitem{46} R.Egger, H.Grabert, {\em Phys.Rev.B}, {\bf 58}, 10761 (1998).
%main_Lutt_junc_theor
\bibitem{47} C.L.Kane, M.P.A.Fisher, {\em Phys.Rev.B}, {\bf 46}, 15233 (1992).
%main_FCS
\bibitem{48} L.S.Levitov, M.Reznikov, {\em Phys.Rev.B}, {\bf 70}, 115305 (2004).
%FCS_theor
\bibitem{49} D.Bagrets, Y.Utsumi, D.Golubev, G.Schön, {\em Fortschr.Phys.}, {\bf 54}, 917 (2006); E.V.Sukhorukov, D.Loss, {\em Electronic correlations: from meso- to nano-physics: proc. of the XXXVIth Rencontres de Moriond},  Les Ulis, pp. 413-417 (2001).

%ours)

\bibitem{50} G.Skorobagatko, A.Bruch, S.V.Kusminskiy, A.Romito, {\em Phys.Rev.B}, {\bf 95}, 205402 (2017).

%orthogonality_cat

%ort_cat
\bibitem{51} P.W.Anderson, {\em Phys.Rev.Lett.} {\bf 18}, 1049 (1967); K.D.Schotte, U.Schotte, {\em Phys.Rev.}, {\bf 182}, 479 (1969).
\bibitem{52} These results will be reported elsewhere.
%main_gen_theor
\bibitem{53} T.Giamarchi, {\em Quantum Physics in One Dimension}, Oxford University Press, Oxford, (2004).
%ort_cat
\bibitem{54} A.Furusaki, {\em Phys.Rev.B}, {\bf 57}, 7141 (1998).

%Kondo_resonance_analogy

%Kondo_res_QEOMS
\bibitem{55} N.Nagaosa, {\em Quantum Field Theory in Strongly Correlated Electronic Systems}, Springer-Verlag Berlin Heidelberg (1999).
%kondo_res
\bibitem{56} M.Pustilnik, L.I.Glazman, {\em J.Phys.Condens.Matter}, {\bf 16}, R513 (2004); G.D.Scott, D.Natelson, {\em Kondo Resonances in Molecular Devices}, ACS Nano, {\bf 4(7)}, 3560–3579 (2010).
\bibitem{57} M.Goldstein, R.Berkovits, Y.Gefen, {\em Phys.Rev.Lett.}, {\bf 104}, 226805 (2010).

\end{references}
\end{document}